\documentclass[%
 reprint,
 superscriptaddress,
 amsmath,amssymb,
 aps,
 pra,
]{revtex4-2}

\usepackage[dvipsnames]{xcolor}
\usepackage{physics}
\usepackage{quantikz}
\usepackage{pgfplots}
\usepackage{float}
\usepackage{graphicx}
\usepackage{dcolumn}
\usepackage{bm}
\usepackage{tikz}
\usetikzlibrary{matrix, decorations.pathreplacing, positioning}

\definecolor{C1}{RGB}{52, 89, 149}
\definecolor{C2}{RGB}{251, 77, 61}
\definecolor{C3}{RGB}{3, 206, 164}
\definecolor{C4}{RGB}{202, 21, 81}
\usepackage{hyperref}
\hypersetup{colorlinks=true, linkcolor=C2, citecolor=C2, urlcolor=C2}

\usepackage[normalem]{ulem}
\usepackage{lipsum}

\newcommand*{\mc}{\mathcal}

\newcommand{\mco}{\mc{O}}

\newcommand{\mbf}{\mathbb{F}}

\newcommand{\mbc}{\mathbb{C}}

\newlength\figureheight
\newlength\figurewidth

\begin{document}

\title{Stabilizer Tensor Networks with Magic State Injection}

\author{Azar C. Nakhl}
\thanks{Contributed Equally}
\affiliation{School of Physics, The University of Melbourne, Parkville, Victoria 3010, Australia}
\author{Ben Harper}
\thanks{Contributed Equally}
\affiliation{School of Physics, The University of Melbourne, Parkville, Victoria 3010, Australia}
\affiliation{Data61, CSIRO, Clayton, Victoria 3168, Australia}
\author{Maxwell West}
\affiliation{School of Physics, The University of Melbourne, Parkville, Victoria 3010, Australia}
\author{Neil Dowling}
\affiliation{School of Physics, The University of Melbourne, Parkville, Victoria 3010, Australia}
\affiliation{School of Physics \& Astronomy, Monash University, Clayton, VIC 3800, Australia}
\author{Martin Sevior}
\affiliation{School of Physics, The University of Melbourne, Parkville, Victoria 3010, Australia}
\author{Thomas Quella}
\affiliation{School of Mathematics and Statistics, The University of Melbourne, Parkville, Victoria 3010, Australia}
\author{Muhammad Usman}
\affiliation{School of Physics, The University of Melbourne, Parkville, Victoria 3010, Australia}
\affiliation{Data61, CSIRO, Clayton, Victoria 3168, Australia}
\affiliation{School of Physics \& Astronomy, Monash University, Clayton, VIC 3800, Australia}

\date{\today}

\begin{abstract}
    This work augments the recently introduced Stabilizer Tensor Network (STN) protocol with magic state injection, reporting a new framework with significantly enhanced ability to simulate circuits with an extensive number of non-Clifford operations. Specifically, for random $T$-doped $N$-qubit Clifford circuits the computational cost of circuits prepared with magic state injection scales as $\mathcal{O}(\text{poly}(N))$ when the circuit has $t \lesssim N$ $T$-gates compared to an exponential scaling for the STN approach, which is demonstrated in systems of up to $200$ qubits. In the case of the Hidden Bit Shift circuit, a paradigmatic benchmarking system for extended stabilizer methods with a tunable amount of magic, we report that our magic state injected STN framework can efficiently simulate $4000$ qubits and $320$ $T$-gates. 
    These findings provide a promising outlook for the use of this protocol in the classical modelling of quantum circuits that are conventionally difficult to simulate efficiently. 
\end{abstract}

\maketitle

\textit{Introduction.---} Advances in the classical simulation of quantum systems are of utmost importance as they underpin the  benchmarking of quantum algorithms~\cite{dang_optimising_2019,west2023provably,PhysRevResearch.5.023186, chen_quantum_2023,nakhl_calibrating_2024,niedermeier_simulating_2024} and enable rigorous tests of nascent claims reporting quantum advantage~\cite{bravyi2019simulation,goh2023lie,kim2023evidence,tindall2024efficient,angrisani2024classically}. 
The state-of-the-art quantum simulation methods primarily revolve around two seemingly independent paradigms: stabilizer methods that can efficiently simulate systems with high entanglement, but struggle with non-Clifford gates~\cite{gottesman1998heisenberg,bravyi2016improved,gidney2021stim,PRXQuantum.3.020361}; or alternatively tensor network-based methods that are well suited to simulations with large number of non-Clifford gates, but scale poorly with entanglement~\cite{schollwock2011density,xu_herculean_2023,nakhl_calibrating_2024}.
Therefore, naturally one may seek to amalgamate these methods in such a way as to harness the advantages of both. Indeed, there has been a significant push towards developing such hybrid simulation approaches ~\cite{lami_quantum_2024,lami_learning_2024,mello_hybrid_2024,qian_clifford_2024,PhysRevB.110.045101,PhysRevLett.133.010601,dowling2024magic,paviglianiti_estimating_2024,qian_augmenting_2024,huang_non-stabilizerness_2024,masot-llima_stabilizer_2024}, with recent work on the Stabilizer Tensor Network (STN) protocol~\cite{masot-llima_stabilizer_2024} presenting an efficient technique for the simulation of both highly-magical systems with low entanglement, and highly entangled systems with low magic, loosely defined by the number of non-Clifford operations in the simulation. Building on this, our work demonstrates a marked improvement in the simulability of highly entangled circuits with a large number of $T$-gates, achieved by incorporating a tool commonly used in quantum error correction, namely magic state injection~\cite{fowler_low_2019}. This is a natural next step for the STN formalism, as several simulation techniques~\cite{bravyi2019simulation,Reardon_Smith_2024} have previously found an advantage by separating computation into special resource states and operations that can be simulated efficiently. Furthermore, our work provides the first demonstrations of the STN protocol being utilised for benchmarking circuits of interest, namely $T$-doped Clifford circuits and the Hidden Bit Shift circuit.

By introducing the use of magic state injection to STNs, we have developed the \textit{Magic state injection Augmented Stabilizer Tensor network} (MAST) simulation framework. To highlight the computational capabilities of the MAST technique, we benchmark MAST and STN against two commonly used circuits, namely $T$-doped Clifford circuits~\cite{bravyi2016improved,mello_hybrid_2024,true_transitions_2022,leone_learning_2024,PRXQuantum.3.020361} and the Hidden Bit Shift circuit~\cite{roetteler_quantum_2009,bravyi2016improved,bravyi2019simulation,kissinger_classical_2022,kissinger_simulating_2022,PRXQuantum.3.020361}. A remarkable finding from our work is that MAST is capable of offering a considerable improvement in simulation cost for some classes of circuits. In particular, for random $T$-doped $N$-qubit Clifford circuits, MAST is able to efficiently simulate circuits with up to $N$ $T$-gates in polynomial time. Beyond this regime, we see a sharp phase transition in simulation cost as the $T$-gate density is scaled. Even in this intermediate-depth case, circuits prepared with magic state injection continue to outperform both standard STN and conventional Matrix Product State (MPS) methods until a saturation depth is reached. For the Hidden Bit Shift circuit we find that MAST outperforms STN for relatively few multiply controlled $Z$ (CCZ) gates. Of note however, is that for both MAST and STN, the bond-dimension appears to scale sub-exponentially, which is in stark contrast to both standard MPS methods and analogous resource costs in stabilizer simulators~\cite{bravyi2016improved,bravyi2019simulation}. Overall, we find that combining STNs with magic state injection is a powerful tool for simulating systems with moderate non-stabilizerness yet high entanglement. 

\textit{Stabilizer Tensor Networks.---} The STN protocol~\cite{masot-llima_stabilizer_2024} is a hybrid simulation method where a quantum state $\ket{\psi}$ is represented using an
MPS in the stabilizer basis as,
\begin{equation}
    \ket{\psi} = \sum_{i=1}^{2^N} \nu_i D_{\hat{\imath}} \ket{\phi}
\end{equation}
where $\nu_i$ are the coefficients of the basis states, $\ket{\phi}$ is a stabilizer state as defined in Appendix~\ref{appendix:prelim} and the $D_{\hat{\imath}}$ are operators defined with respect to the destabilizer tableau.

In this protocol, Clifford operations (the gate set generated by $\{ H, S, CNOT \}$ gates) correspond to an update of the basis that can be kept track of in a regular stabilizer tableau~\cite{aaronson_improved_2008}, whilst non-Clifford operations can be represented as a sum over elements of the stabilizer basis and correspond to a controlled rotation on the MPS. Expectation values for operators in the stabilizer basis are found using standard MPS methods, which is in general efficient for an MPS in canonical form~\cite{schollwock2011density}. Details on the implementation of STN can be found in the Appendix~\ref{appendix:stn}.

An interesting consequence of the set-up of the STN protocol is that non-Clifford operations on un-entangled qubits may be performed in polynomial-time. In contrast, projective measurement, which in tensor-network and stabilizer simulations is computationally efficient, is in fact computationally complex in the STN protocol, corresponding to a sequence of entangling gates on the tensor-network. This can play a significant role in the simulation cost of some classes of circuit.

\begin{figure}
    \centering
    \includegraphics{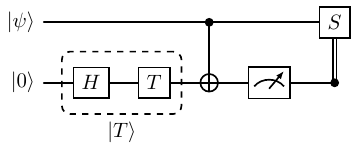}
    \caption{Magic state injection consists of local operations preparing a magic state, then injection using Clifford gates. If the magic state is measured to be $\ket{1}$, a correction gate must be applied. This figure shows the magic state gadget for applying a $T$-gate (a $\pi/4$ rotation around $Z$). In this case, the correction is a Clifford gate, but this is not the case in general.}
    \label{fig:state-injection}
\end{figure}

\textit{Magic State Injected Stabilizer Tensor Networks.---}
Building on the observation above, we note that in an arbitrary quantum circuit, non-Clifford operations may be replaced by a \textit{state injection gadget}, shown in Figure~\ref{fig:state-injection}. We improve STN with the addition of magic state injection, referred to as MAST. When using MAST, magic state preparation and simulation of the resulting Clifford circuit may be performed in polynomial time, noting that in simulation, it is possible to predetermine the outcome of the measurement in the state injection gadget and delay the actual projection until the end of the circuit. The primary cost of a simulation is hence delayed until the final projection step.

It is not immediately obvious that this reframing leads to a computational advantage --- in fact, it adds more ancilla qubits and measurements to achieve the same calculation.
However, it is thought that the stabilizer tableau in the STN framework stores ``potential entanglement'' that only transfers to the tensor network when a non-Clifford operation is applied~\cite{masot-llima_stabilizer_2024}. 
We posit that an arbitrary circuit may result in an intermediate state with more entanglement than the final state requires~\cite{dupont_entanglement_2022}. In this case, the projection at the end may be cheaper than applying $T$-gates directly. On the other hand, as the cost of simulation scales exponentially with the magic required, the additive computational cost of adding extra ancillas is negligible.

\begin{figure*}
    \includegraphics[width=\textwidth]{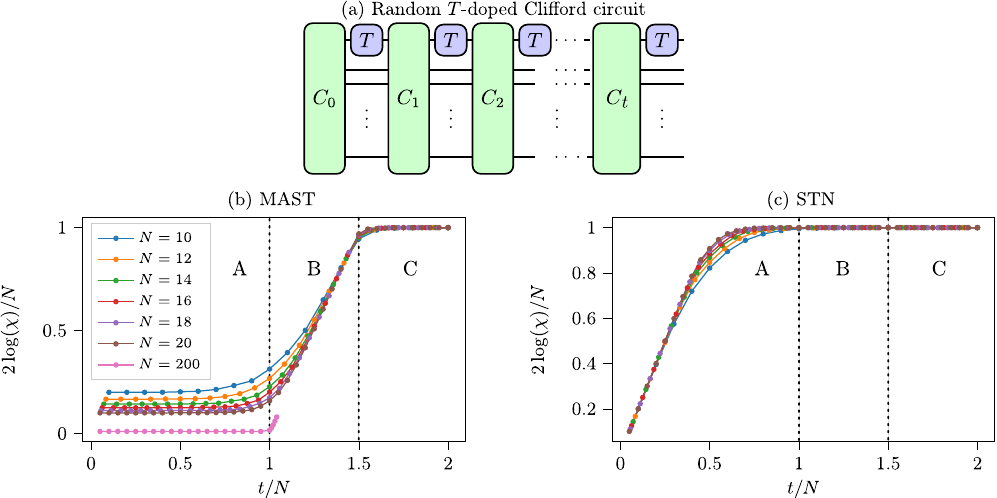}
    \caption{(a) A schematic illustration of random $T$-doped Clifford circuits. The Clifford gates $C_i$ are generated using the Qiskit SDK~\cite{qiskit} using the procedure described in Ref.~\cite{bravyi2021hadamard}. (b)-(c)  Simulations with the MAST protocol (b) and STN protocol (c) averaged over $1000$ random instances. We show the maximum 
    bond-dimension $\chi$ per MPS required for the complete simulation of a circuit with $t$ $T$-gates. Both these axes are scaled with respect to the number of data qubits (i.e. not including the ancilla space for the MAST protocol). We recognise three distinct regions; Region A for $t\lesssim N$ where MAST has an average bond-dimension bounded by $3$ and the bond-dimension for STN exponentially increases with increasing $t$; Region B for $N \lesssim t \lesssim 1.5N$ where the bond-dimension for MAST exponentially increases with increasing $t$ and where STN has the maximal bond dimension $\chi=2^{N/2}$; Region C where both MAST and STN have the maximal bond-dimension of $\chi=2^{N/2}$. 
    We do not simulate more than $N+10$ layers for $N=200$ due to the exponential increase in bond-dimension.
    }
    \label{fig:rand}
\end{figure*}

\textit{Results and Discussion.---}
To benchmark the effectiveness of both standard and magic state injected STNs, we perform simulations of random $T$-doped Clifford circuits as well as the hidden shift circuit. 
Additionally, to test whether eliminating the cost of the potential entanglement in the stabilizer tableau of the STN protocol is the source of advantage in MAST compared to STN~\cite{masot-llima_stabilizer_2024}, we run the circuit $U^\dagger U\ket{0}$ for $U$ generated as a sequence of Clifford + $T$-gates as above. While trivial in principle, this computation is in general difficult for standard MPS and stabilizer simulation as a result of the intermediate entanglement and magic generated in the circuit. 

\textit{A. Random Circuits.---}
We first consider $T$-doped Clifford circuits~\cite{bravyi2021hadamard} of varying depths, as depicted schematically in Figure \ref{fig:rand}(a). We define a single layer of such a circuit to consist  of a uniformly randomly drawn $N$-qubit Clifford operation followed by a $T$-gate on the first qubit (the left-right invariance of the uniform measure on the Cliffords combined with the well-known fact that the qubit-permuting operators are Clifford implies that this is equivalent to randomly selecting the $T$-gate position at each layer). 
For depth (number of layers) $t\gg \log(N)$
this ensemble of random circuits is intractable to simulate using both standard tensor network and stabilizer methods, due respectively to the high entanglement generation of random Cliffords and the exponential-in-$t$ scaling of magic~\cite{aaronson2004improved,eisert_entanglement_2013}, and therefore constitutes an ideal testbed for MAST.

In Appendix~\ref{app:proof} we show that on average, MAST can simulate random $T$-doped Clifford circuits efficiently when $t$, the number of $T$ gates, is less than $N$, the number of qubits. The introduction of MAST is key to this argument, as it allows us to make assumptions about the structure of the constituent stabilizer tableau and tensor network in the simulator. Briefly, the argument is as follows; the cost of simulation in MAST comes from the projective measurement of magic states, which corresponds to an entangling operation in the MPS. We show that if this operation is on a certain subset of the MPS, it is not in fact entangling and does not increase the bond dimension. Finally, we show that for random $T$-doped Cliffords, there is a high probability of this being true when $t \lesssim N$.

In Figure \ref{fig:rand}(b)-(c) we present numerical simulation results comparing MAST to STN, and show that we are able to simulate up to $N$ layers with bounded average bond dimension using MAST (Region A), in contrast to the  exponentially scaling bond dimension displayed by STNs. 
In the intermediate-depth case of $N\lesssim t\lesssim 1.5 N$ (Region B) we observe a ``phase transition'', with the MAST bond dimension increasing exponentially in the number of additional layers. Despite the increase in computational resources required, MAST continues to significantly outperform STNs throughout Region B. 
Finally, for depths $t\gtrsim 1.5N$ (Region C), the MAST bond dimension reaches the maximum value of $2^{N/2}$.

Now, as is well known, the resources required to simulate an $N$-qubit Clifford circuit doped with $t$ $T$-gates via the standard stabiliser framework scale as $\mco({\rm poly}(N)\exp (t))$ and become intractably large when $t=\mco(N)$~\cite{gottesman1998heisenberg,bravyi2016improved}. More generally, it is strongly believed that the simulation of Clifford circuits with $\mathcal{O(N)}$ non-Clifford resources is generically intractable~\cite{bremner2011classical}.
In this regime, for example, it has been shown that doped Clifford circuits form approximate unitary designs to a precision exponentially small in the system size, reproducing long-time signatures of quantum chaos~\cite{Leone2021quantumchaosis}. 
In contrast,
we see that MAST is consuming only bounded classical resources up until a critical threshold of $t \approx N$, which is exactly the expected onset of intractable quantum phenomena~\cite{bremner2011classical}. We also note that in the regime $t\leq N$, previous work has shown surprising advantages in unitary learning~\cite{PhysRevA.109.022429,PhysRevLett.132.080402} and entanglement manipulation~\cite{gu2024magicinducedcomp}. Our results therefore adds to the growing list of useful properties of this `entanglement dominated' phase.
Noting that MAST allows for efficient computation of local expectations but does not admit efficient sampling from the probability distributions, our results appear commensurate with recent works~\cite{mele2024noise,angrisani2024classically,PRXQuantum.3.020361} that show the efficient simulability of Pauli expectation values from ``locally-scrambling'' families of circuits. In particular, we note that while the stabilizer-based technique in~\cite{PRXQuantum.3.020361} sees similar performance on this class of circuit and deterministic quantum circuits, it makes no improvement to the general problem of sampling from the probability distribution of a state, which scales as $\mathcal{O}(\exp(t) N^3 w^3)$, where $w$ is the number of bits being sampled (in the case of deterministic circuits we take $w=1$ for each qubit). In contrast, MAST scales as $\mathcal{O}(\exp(w))$ when $t < N$, so in the case where $t \gg w$, MAST will be more efficient at this computationally harder problem. In circuits where the final state has low entanglement, such as the Hidden Bit Shift below, sampling from the probability distribution will not increase the bond dimension and hence simulation cost, potentially giving MAST a large advantage over other techniques.

\begin{figure}
    \hspace{-5mm}
    \includegraphics[]{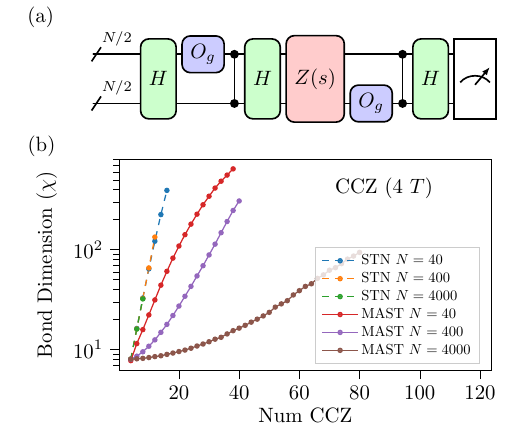}
    \caption{(a) Hidden Bit Shift Circuit. $O_g$ is a random phase gate constructed from $\{Z, CZ, CCZ\}$. $Z(s)$ encodes the hidden bit shift, and the final measurement will result in $s$ with 100\% probability. (b) Simulation cost of the Hidden Bit Shift circuit for up to $4000$ qubits. The number of Clifford operations in $O_g$ is scaled proportionately with $N$ to ensure sufficient mixing. In these simulations, the CCZ gate is decomposed into 4 $T$ gates and an ancilla as per Appendix~\ref{apn:ccz}. Appendix~\ref{app:hidden-shift} features results and discussion for an alternative decomposition of the CCZ gates. As in Figure~\ref{fig:rand}, increasing the number of qubits for a fixed number of $T$-gates reduces the simulation cost.}\label{fig:hidden-shift}
\end{figure}

In addition to small system sizes of fewer than $20$ qubits, we also simulate up to $t=N+10$ layers for $N=200$ qubits demonstrating that this simulation protocol is scalable. Finally, though not included in Figure~\ref{fig:rand}, we also simulated random circuits with arbitrary $R_z(\theta)$ rotations replacing the $T$-gates, resulting in an identical distribution for the bond-dimension. This is unsurprising, as the different angle only changes the coefficients of the gate decomposition in the STN framework (see Appendix~\ref{appendix:stn_non}), which ultimately does not influence simulation cost.

Finally, we consider the task of numerically simulating the trivial circuit $U^\dagger U \ket{0}$.
We find that for these simulations the bond-dimension is in general significantly greater than $1$, indicating that MAST does not universally determine the lowest possible resource cost required for any circuit. We do, however, observe that the resource cost is path dependent where one is free to choose the order in which ancilla qubits are projected. For the particular case of random circuits, projecting pairwise from the middle of the ancilla register out towards either end of that register, one finds that the bond-dimension of the MPS is at most $2$ throughout the simulation. This is as a result of each pair of projections generating two CNOT cascades as per Appendix~\ref{apn:stn_proj} that cancel each other out such that the result after the pair of projections is a product state.

\textit{B. Hidden Bit Shift.---}
We now consider the Hidden Bit Shift circuit~\cite{roetteler_quantum_2009}. This circuit has two desirable properties for benchmarking; first, it creates a large amount of entanglement, making it difficult to implement for all but stabilizer simulators and fault-tolerant quantum computers, and secondly, a tuneable number of non-Clifford operations, allowing one to easily analyse the performance of a simulator for different amounts of magic. The class of circuits simulated in this work is identical to that of Refs.~\cite{bravyi2016improved,bravyi2019simulation}, and is shown in Figure~\ref{fig:hidden-shift}(a). 

Compared to Ref.~\cite{bravyi2019simulation}, which simulates up to 64 $T$-gates on $40$ qubits using an extended stabilizer simulator that scales exponentially with the number of $T$-gates, Figure~\ref{fig:hidden-shift}(b) shows that our technique is able to simulate 160 $T$-gates on $40$ qubits. In addition to this, we increase the size of the circuits being simulated. As per the results for random $T$-doped Clifford circuits, we find that increasing the number of qubits in the Hidden Bit Shift circuit does indeed reduce the bond-dimension under MAST, allowing large circuits of $4000$ qubits and $320$ $T$-gates to be simulated on a classical computer. Conversely, STN performs worse at higher qubit counts. Finally, we note that it has recently been shown theoretically that the hidden-shift circuit can be simulated in polynomial-time~\cite{amy_polynomial-time_2024}, although this has not been demonstrated in simulation.

\textit{Conclusion and Outlook.---}
The Magic state injection Augmented Stabilizer Tensor network (MAST) method is capable of simulating large circuits that are both highly entangled and magical. In particular, we highlight that random $T$-doped Clifford circuits can be efficiently simulated to depths less than the number of qubits and show numerical results with $\mathcal{O}(N)$ $T$-gates for up to $N=200$ qubits with bounded STN bond dimension. Recently, a similar observation was made for the Clifford Augmented Matrix Product State (CAMPS) protocol~\cite{fux_disentangling_2024}, which is a different amalgamation of stabilizer and tensor network methods~\cite{lami_quantum_2024,lami_quantum_2024,qian_augmenting_2024,huang_non-stabilizerness_2024} to the STNs considered in this work. Unlike the simulation method devised in this work, the CAMPS-based method requires an optimisation subroutine whereas our method does not.

The MAST simulation protocol introduced in this work, and stabilizer tensor network methods more broadly have wide reaching applicability and may be used to simulate a variety of quantum algorithms with fewer computational resources than current state-of-the-art simulation protocols. There are many algorithms of interest to the research community with only a moderate amount of entanglement and magic, in particular Noisy Intermediate Scale Quantum (NISQ)~\cite{preskill2018quantum} algorithms such as the Quantum Approximate Optimisation Algorithm (QAOA)~\cite{farhi2014quantum}, the Variational Quantum Eigensolver (VQE)~\cite{peruzzo2014variational}, or Quantum Machine Learning (QML)~\cite{west2023provably}. Our work may allow larger simulations of these algorithms using classical resources.

In addition, there is room to further refine MAST and STNs as simulation techniques. As shown in this work, the simulation cost is highly dependent on the decomposition of expensive operations and it may be possible to further optimise these decompositions, similar to the work done optimising magic state decompositions for stabilizer simulators~\cite{bravyi2019simulation,Qassim2021improvedupperbounds}. This also leads to a theoretical understanding relating the simulation cost of MAST to a property of the quantum state being simulated, similar to bond-dimension and entanglement in a conventional tensor network, or stabilizer rank and magic. This would not only allow for a more direct comparison with other simulation methods where the resource cost is well established but it would also shed a light on precisely what makes certain quantum systems difficult to simulate and narrow down the regime where quantum computers have an advantage over classical computers.

\textit{Acknowledgements.---} A.C.N., M.W., and N.D. acknowledge the support of Australian Government Research Training Program Scholarships. B.H. acknowledges the CSIRO Research Training Program Scholarship. Computational resources were provided by the Pawsey Supercomputing Research Center through the National Computational Merit Allocation Scheme (NCMAS). The research was supported by the University of Melbourne through the establishment of the IBM Quantum Network Hub at the University.
\\
Data Availability: The implementation of the MAST protocol used to perform all numerical simulations featured in this work can be found in~\cite{mastgit}. The data generated by the numerical simulations can be found in the figures.
\\ \\ 
\noindent
{\small \phantom{...} email: anakhl@student.unimelb.edu.au }\\
\noindent
{\small \phantom{...} email: bharper1@student.unimelb.edu.au }\\

\bibliography{referenceFree} 

\begin{thebibliography}{58}%
\makeatletter
\providecommand \@ifxundefined [1]{%
 \@ifx{#1\undefined}
}%
\providecommand \@ifnum [1]{%
 \ifnum #1\expandafter \@firstoftwo
 \else \expandafter \@secondoftwo
 \fi
}%
\providecommand \@ifx [1]{%
 \ifx #1\expandafter \@firstoftwo
 \else \expandafter \@secondoftwo
 \fi
}%
\providecommand \natexlab [1]{#1}%
\providecommand \enquote  [1]{``#1''}%
\providecommand \bibnamefont  [1]{#1}%
\providecommand \bibfnamefont [1]{#1}%
\providecommand \citenamefont [1]{#1}%
\providecommand \href@noop [0]{\@secondoftwo}%
\providecommand \href [0]{\begingroup \@sanitize@url \@href}%
\providecommand \@href[1]{\@@startlink{#1}\@@href}%
\providecommand \@@href[1]{\endgroup#1\@@endlink}%
\providecommand \@sanitize@url [0]{\catcode `\\12\catcode `\$12\catcode
  `\&12\catcode `\#12\catcode `\^12\catcode `\_12\catcode `\%12\relax}%
\providecommand \@@startlink[1]{}%
\providecommand \@@endlink[0]{}%
\providecommand \url  [0]{\begingroup\@sanitize@url \@url }%
\providecommand \@url [1]{\endgroup\@href {#1}{\urlprefix }}%
\providecommand \urlprefix  [0]{URL }%
\providecommand \Eprint [0]{\href }%
\providecommand \doibase [0]{https://doi.org/}%
\providecommand \selectlanguage [0]{\@gobble}%
\providecommand \bibinfo  [0]{\@secondoftwo}%
\providecommand \bibfield  [0]{\@secondoftwo}%
\providecommand \translation [1]{[#1]}%
\providecommand \BibitemOpen [0]{}%
\providecommand \bibitemStop [0]{}%
\providecommand \bibitemNoStop [0]{.\EOS\space}%
\providecommand \EOS [0]{\spacefactor3000\relax}%
\providecommand \BibitemShut  [1]{\csname bibitem#1\endcsname}%
\let\auto@bib@innerbib\@empty
\bibitem [{\citenamefont {Dang}\ \emph {et~al.}(2019)\citenamefont {Dang},
  \citenamefont {Hill},\ and\ \citenamefont
  {Hollenberg}}]{dang_optimising_2019}%
  \BibitemOpen
  \bibfield  {author} {\bibinfo {author} {\bibfnamefont {A.}~\bibnamefont
  {Dang}}, \bibinfo {author} {\bibfnamefont {C.~D.}\ \bibnamefont {Hill}},\
  and\ \bibinfo {author} {\bibfnamefont {L.~C.~L.}\ \bibnamefont
  {Hollenberg}},\ }\bibfield  {title} {\bibinfo {title} {Optimising {Matrix}
  {Product} {State} {Simulations} of {Shor}'s {Algorithm}},\ }\href
  {https://doi.org/10.22331/q-2019-01-25-116} {\bibfield  {journal} {\bibinfo
  {journal} {Quantum}\ }\textbf {\bibinfo {volume} {3}},\ \bibinfo {pages}
  {116} (\bibinfo {year} {2019})},\ \bibinfo {note} {arXiv:1712.07311
  [quant-ph]}\BibitemShut {NoStop}%
\bibitem [{\citenamefont {West}\ \emph {et~al.}(2024)\citenamefont {West},
  \citenamefont {Heredge}, \citenamefont {Sevior},\ and\ \citenamefont
  {Usman}}]{west2023provably}%
  \BibitemOpen
  \bibfield  {author} {\bibinfo {author} {\bibfnamefont {M.~T.}\ \bibnamefont
  {West}}, \bibinfo {author} {\bibfnamefont {J.}~\bibnamefont {Heredge}},
  \bibinfo {author} {\bibfnamefont {M.}~\bibnamefont {Sevior}},\ and\ \bibinfo
  {author} {\bibfnamefont {M.}~\bibnamefont {Usman}},\ }\bibfield  {title}
  {\bibinfo {title} {{Provably} {Trainable} {Rotationally} {Equivariant}
  {Quantum} {Machine} {Learning}},\ }\href
  {https://doi.org/10.1103/PRXQuantum.5.030320} {\bibfield  {journal} {\bibinfo
   {journal} {PRX Quantum}\ }\textbf {\bibinfo {volume} {5}},\ \bibinfo {pages}
  {030320} (\bibinfo {year} {2024})}\BibitemShut {NoStop}%
\bibitem [{\citenamefont {West}\ \emph {et~al.}(2023)\citenamefont {West},
  \citenamefont {Erfani}, \citenamefont {Leckie}, \citenamefont {Sevior},
  \citenamefont {Hollenberg},\ and\ \citenamefont
  {Usman}}]{PhysRevResearch.5.023186}%
  \BibitemOpen
  \bibfield  {author} {\bibinfo {author} {\bibfnamefont {M.~T.}\ \bibnamefont
  {West}}, \bibinfo {author} {\bibfnamefont {S.~M.}\ \bibnamefont {Erfani}},
  \bibinfo {author} {\bibfnamefont {C.}~\bibnamefont {Leckie}}, \bibinfo
  {author} {\bibfnamefont {M.}~\bibnamefont {Sevior}}, \bibinfo {author}
  {\bibfnamefont {L.~C.~L.}\ \bibnamefont {Hollenberg}},\ and\ \bibinfo
  {author} {\bibfnamefont {M.}~\bibnamefont {Usman}},\ }\bibfield  {title}
  {\bibinfo {title} {Benchmarking adversarially robust quantum machine learning
  at scale},\ }\href {https://doi.org/10.1103/PhysRevResearch.5.023186}
  {\bibfield  {journal} {\bibinfo  {journal} {Phys. Rev. Res.}\ }\textbf
  {\bibinfo {volume} {5}},\ \bibinfo {pages} {023186} (\bibinfo {year}
  {2023})}\BibitemShut {NoStop}%
\bibitem [{\citenamefont {Chen}\ \emph {et~al.}(2023)\citenamefont {Chen},
  \citenamefont {Stoudenmire},\ and\ \citenamefont
  {White}}]{chen_quantum_2023}%
  \BibitemOpen
  \bibfield  {author} {\bibinfo {author} {\bibfnamefont {J.}~\bibnamefont
  {Chen}}, \bibinfo {author} {\bibfnamefont {E.}~\bibnamefont {Stoudenmire}},\
  and\ \bibinfo {author} {\bibfnamefont {S.~R.}\ \bibnamefont {White}},\
  }\bibfield  {title} {\bibinfo {title} {Quantum {Fourier} {Transform} has
  {Small} {Entanglement}},\ }\href
  {https://doi.org/10.1103/PRXQuantum.4.040318} {\bibfield  {journal} {\bibinfo
   {journal} {PRX Quantum}\ }\textbf {\bibinfo {volume} {4}},\ \bibinfo {pages}
  {040318} (\bibinfo {year} {2023})}\BibitemShut {NoStop}%
\bibitem [{\citenamefont {Nakhl}\ \emph {et~al.}(2024)\citenamefont {Nakhl},
  \citenamefont {Quella},\ and\ \citenamefont
  {Usman}}]{nakhl_calibrating_2024}%
  \BibitemOpen
  \bibfield  {author} {\bibinfo {author} {\bibfnamefont {A.~C.}\ \bibnamefont
  {Nakhl}}, \bibinfo {author} {\bibfnamefont {T.}~\bibnamefont {Quella}},\ and\
  \bibinfo {author} {\bibfnamefont {M.}~\bibnamefont {Usman}},\ }\bibfield
  {title} {\bibinfo {title} {Calibrating the role of entanglement in
  variational quantum circuits},\ }\href
  {https://doi.org/10.1103/PhysRevA.109.032413} {\bibfield  {journal} {\bibinfo
   {journal} {Physical Review A}\ }\textbf {\bibinfo {volume} {109}},\ \bibinfo
  {pages} {032413} (\bibinfo {year} {2024})}\BibitemShut {NoStop}%
\bibitem [{\citenamefont {Niedermeier}\ \emph {et~al.}(2024)\citenamefont
  {Niedermeier}, \citenamefont {Lado},\ and\ \citenamefont
  {Flindt}}]{niedermeier_simulating_2024}%
  \BibitemOpen
  \bibfield  {author} {\bibinfo {author} {\bibfnamefont {M.}~\bibnamefont
  {Niedermeier}}, \bibinfo {author} {\bibfnamefont {J.~L.}\ \bibnamefont
  {Lado}},\ and\ \bibinfo {author} {\bibfnamefont {C.}~\bibnamefont {Flindt}},\
  }\bibfield  {title} {\bibinfo {title} {Simulating the quantum {Fourier}
  transform, {Grover}'s algorithm, and the quantum counting algorithm with
  limited entanglement using tensor networks},\ }\href
  {https://doi.org/10.1103/PhysRevResearch.6.033325} {\bibfield  {journal}
  {\bibinfo  {journal} {Physical Review Research}\ }\textbf {\bibinfo {volume}
  {6}},\ \bibinfo {pages} {033325} (\bibinfo {year} {2024})}\BibitemShut
  {NoStop}%
\bibitem [{\citenamefont {Bravyi}\ \emph {et~al.}(2019)\citenamefont {Bravyi},
  \citenamefont {Browne}, \citenamefont {Calpin}, \citenamefont {Campbell},
  \citenamefont {Gosset},\ and\ \citenamefont {Howard}}]{bravyi2019simulation}%
  \BibitemOpen
  \bibfield  {author} {\bibinfo {author} {\bibfnamefont {S.}~\bibnamefont
  {Bravyi}}, \bibinfo {author} {\bibfnamefont {D.}~\bibnamefont {Browne}},
  \bibinfo {author} {\bibfnamefont {P.}~\bibnamefont {Calpin}}, \bibinfo
  {author} {\bibfnamefont {E.}~\bibnamefont {Campbell}}, \bibinfo {author}
  {\bibfnamefont {D.}~\bibnamefont {Gosset}},\ and\ \bibinfo {author}
  {\bibfnamefont {M.}~\bibnamefont {Howard}},\ }\bibfield  {title} {\bibinfo
  {title} {Simulation of quantum circuits by low-rank stabilizer
  decompositions},\ }\href {https://doi.org/10.22331/q-2019-09-02-181}
  {\bibfield  {journal} {\bibinfo  {journal} {Quantum}\ }\textbf {\bibinfo
  {volume} {3}},\ \bibinfo {pages} {181} (\bibinfo {year} {2019})}\BibitemShut
  {NoStop}%
\bibitem [{\citenamefont {Goh}\ \emph {et~al.}(2023)\citenamefont {Goh},
  \citenamefont {Larocca}, \citenamefont {Cincio}, \citenamefont {Cerezo},\
  and\ \citenamefont {Sauvage}}]{goh2023lie}%
  \BibitemOpen
  \bibfield  {author} {\bibinfo {author} {\bibfnamefont {M.~L.}\ \bibnamefont
  {Goh}}, \bibinfo {author} {\bibfnamefont {M.}~\bibnamefont {Larocca}},
  \bibinfo {author} {\bibfnamefont {L.}~\bibnamefont {Cincio}}, \bibinfo
  {author} {\bibfnamefont {M.}~\bibnamefont {Cerezo}},\ and\ \bibinfo {author}
  {\bibfnamefont {F.}~\bibnamefont {Sauvage}},\ }\bibfield  {title} {\bibinfo
  {title} {Lie-algebraic classical simulations for variational quantum
  computing},\ }\href {https://arxiv.org/abs/2308.01432} {\bibfield  {journal}
  {\bibinfo  {journal} {arXiv preprint arXiv:2308.01432}\ } (\bibinfo {year}
  {2023})}\BibitemShut {NoStop}%
\bibitem [{\citenamefont {Kim}\ \emph {et~al.}(2023)\citenamefont {Kim},
  \citenamefont {Eddins}, \citenamefont {Anand}, \citenamefont {Wei},
  \citenamefont {Van Den~Berg}, \citenamefont {Rosenblatt}, \citenamefont
  {Nayfeh}, \citenamefont {Wu}, \citenamefont {Zaletel}, \citenamefont {Temme}
  \emph {et~al.}}]{kim2023evidence}%
  \BibitemOpen
  \bibfield  {author} {\bibinfo {author} {\bibfnamefont {Y.}~\bibnamefont
  {Kim}}, \bibinfo {author} {\bibfnamefont {A.}~\bibnamefont {Eddins}},
  \bibinfo {author} {\bibfnamefont {S.}~\bibnamefont {Anand}}, \bibinfo
  {author} {\bibfnamefont {K.~X.}\ \bibnamefont {Wei}}, \bibinfo {author}
  {\bibfnamefont {E.}~\bibnamefont {Van Den~Berg}}, \bibinfo {author}
  {\bibfnamefont {S.}~\bibnamefont {Rosenblatt}}, \bibinfo {author}
  {\bibfnamefont {H.}~\bibnamefont {Nayfeh}}, \bibinfo {author} {\bibfnamefont
  {Y.}~\bibnamefont {Wu}}, \bibinfo {author} {\bibfnamefont {M.}~\bibnamefont
  {Zaletel}}, \bibinfo {author} {\bibfnamefont {K.}~\bibnamefont {Temme}},
  \emph {et~al.},\ }\bibfield  {title} {\bibinfo {title} {Evidence for the
  utility of quantum computing before fault tolerance},\ }\href
  {https://doi.org/10.1038/s41586-023-06096-3} {\bibfield  {journal} {\bibinfo
  {journal} {Nature}\ }\textbf {\bibinfo {volume} {618}},\ \bibinfo {pages}
  {500} (\bibinfo {year} {2023})}\BibitemShut {NoStop}%
\bibitem [{\citenamefont {Tindall}\ \emph {et~al.}(2024)\citenamefont
  {Tindall}, \citenamefont {Fishman}, \citenamefont {Stoudenmire},\ and\
  \citenamefont {Sels}}]{tindall2024efficient}%
  \BibitemOpen
  \bibfield  {author} {\bibinfo {author} {\bibfnamefont {J.}~\bibnamefont
  {Tindall}}, \bibinfo {author} {\bibfnamefont {M.}~\bibnamefont {Fishman}},
  \bibinfo {author} {\bibfnamefont {E.~M.}\ \bibnamefont {Stoudenmire}},\ and\
  \bibinfo {author} {\bibfnamefont {D.}~\bibnamefont {Sels}},\ }\bibfield
  {title} {\bibinfo {title} {Efficient {Tensor} {Network} {Simulation} of
  {IBM}’s {Eagle} {Kicked} {Ising} {Experiment}},\ }\href
  {https://doi.org/10.1103/PRXQuantum.5.010308} {\bibfield  {journal} {\bibinfo
   {journal} {PRX Quantum}\ }\textbf {\bibinfo {volume} {5}},\ \bibinfo {pages}
  {010308} (\bibinfo {year} {2024})}\BibitemShut {NoStop}%
\bibitem [{\citenamefont {Angrisani}\ \emph {et~al.}(2024)\citenamefont
  {Angrisani}, \citenamefont {Schmidhuber}, \citenamefont {Rudolph},
  \citenamefont {Cerezo}, \citenamefont {Holmes},\ and\ \citenamefont
  {Huang}}]{angrisani2024classically}%
  \BibitemOpen
  \bibfield  {author} {\bibinfo {author} {\bibfnamefont {A.}~\bibnamefont
  {Angrisani}}, \bibinfo {author} {\bibfnamefont {A.}~\bibnamefont
  {Schmidhuber}}, \bibinfo {author} {\bibfnamefont {M.~S.}\ \bibnamefont
  {Rudolph}}, \bibinfo {author} {\bibfnamefont {M.}~\bibnamefont {Cerezo}},
  \bibinfo {author} {\bibfnamefont {Z.}~\bibnamefont {Holmes}},\ and\ \bibinfo
  {author} {\bibfnamefont {H.-Y.}\ \bibnamefont {Huang}},\ }\bibfield  {title}
  {\bibinfo {title} {Classically estimating observables of noiseless quantum
  circuits},\ }\href {https://arxiv.org/abs/2409.01706} {\bibfield  {journal}
  {\bibinfo  {journal} {arXiv preprint arXiv:2409.01706}\ } (\bibinfo {year}
  {2024})}\BibitemShut {NoStop}%
\bibitem [{\citenamefont {Gottesman}(1998)}]{gottesman1998heisenberg}%
  \BibitemOpen
  \bibfield  {author} {\bibinfo {author} {\bibfnamefont {D.}~\bibnamefont
  {Gottesman}},\ }\bibfield  {title} {\bibinfo {title} {The {Heisenberg}
  representation of quantum computers, talk at},\ }in\ \href
  {http://citeseerx.ist.psu.edu/viewdoc/summary?doi=10.1.1.252.9446} {\emph
  {\bibinfo {booktitle} {International Conference on Group Theoretic Methods in
  Physics}}}\ (\bibinfo {organization} {Citeseer},\ \bibinfo {year}
  {1998})\BibitemShut {NoStop}%
\bibitem [{\citenamefont {Bravyi}\ and\ \citenamefont
  {Gosset}(2016)}]{bravyi2016improved}%
  \BibitemOpen
  \bibfield  {author} {\bibinfo {author} {\bibfnamefont {S.}~\bibnamefont
  {Bravyi}}\ and\ \bibinfo {author} {\bibfnamefont {D.}~\bibnamefont
  {Gosset}},\ }\bibfield  {title} {\bibinfo {title} {{Improved} {Classical}
  {Simulation} of {Quantum} {Circuits} {Dominated} by {Clifford} {Gates}},\
  }\href {https://doi.org/10.1103/PhysRevLett.116.250501} {\bibfield  {journal}
  {\bibinfo  {journal} {Phys. Rev. Lett.}\ }\textbf {\bibinfo {volume} {116}},\
  \bibinfo {pages} {250501} (\bibinfo {year} {2016})}\BibitemShut {NoStop}%
\bibitem [{\citenamefont {Gidney}(2021)}]{gidney2021stim}%
  \BibitemOpen
  \bibfield  {author} {\bibinfo {author} {\bibfnamefont {C.}~\bibnamefont
  {Gidney}},\ }\bibfield  {title} {\bibinfo {title} {Stim: a fast stabilizer
  circuit simulator},\ }\href {https://doi.org/10.22331/q-2021-07-06-497}
  {\bibfield  {journal} {\bibinfo  {journal} {Quantum}\ }\textbf {\bibinfo
  {volume} {5}},\ \bibinfo {pages} {497} (\bibinfo {year} {2021})}\BibitemShut
  {NoStop}%
\bibitem [{\citenamefont {Pashayan}\ \emph {et~al.}(2022)\citenamefont
  {Pashayan}, \citenamefont {Reardon-Smith}, \citenamefont {Korzekwa},\ and\
  \citenamefont {Bartlett}}]{PRXQuantum.3.020361}%
  \BibitemOpen
  \bibfield  {author} {\bibinfo {author} {\bibfnamefont {H.}~\bibnamefont
  {Pashayan}}, \bibinfo {author} {\bibfnamefont {O.}~\bibnamefont
  {Reardon-Smith}}, \bibinfo {author} {\bibfnamefont {K.}~\bibnamefont
  {Korzekwa}},\ and\ \bibinfo {author} {\bibfnamefont {S.~D.}\ \bibnamefont
  {Bartlett}},\ }\bibfield  {title} {\bibinfo {title} {Fast estimation of
  outcome probabilities for quantum circuits},\ }\href
  {https://doi.org/10.1103/PRXQuantum.3.020361} {\bibfield  {journal} {\bibinfo
   {journal} {PRX Quantum}\ }\textbf {\bibinfo {volume} {3}},\ \bibinfo {pages}
  {020361} (\bibinfo {year} {2022})}\BibitemShut {NoStop}%
\bibitem [{\citenamefont {Schollw{\"o}ck}(2011)}]{schollwock2011density}%
  \BibitemOpen
  \bibfield  {author} {\bibinfo {author} {\bibfnamefont {U.}~\bibnamefont
  {Schollw{\"o}ck}},\ }\bibfield  {title} {\bibinfo {title} {The density-matrix
  renormalization group in the age of matrix product states},\ }\href
  {https://www.sciencedirect.com/science/article/pii/S0003491610001752?via%3Dihub}
  {\bibfield  {journal} {\bibinfo  {journal} {Annals of physics}\ }\textbf
  {\bibinfo {volume} {326}},\ \bibinfo {pages} {96} (\bibinfo {year}
  {2011})}\BibitemShut {NoStop}%
\bibitem [{\citenamefont {Xu}\ \emph {et~al.}(2023)\citenamefont {Xu},
  \citenamefont {Benjamin}, \citenamefont {Sun}, \citenamefont {Yuan},\ and\
  \citenamefont {Zhang}}]{xu_herculean_2023}%
  \BibitemOpen
  \bibfield  {author} {\bibinfo {author} {\bibfnamefont {X.}~\bibnamefont
  {Xu}}, \bibinfo {author} {\bibfnamefont {S.}~\bibnamefont {Benjamin}},
  \bibinfo {author} {\bibfnamefont {J.}~\bibnamefont {Sun}}, \bibinfo {author}
  {\bibfnamefont {X.}~\bibnamefont {Yuan}},\ and\ \bibinfo {author}
  {\bibfnamefont {P.}~\bibnamefont {Zhang}},\ }\href
  {http://arxiv.org/abs/2302.08880} {\bibinfo {title} {A {Herculean} task:
  {Classical} simulation of quantum computers}} (\bibinfo {year} {2023}),\
  \bibinfo {note} {arXiv:2302.08880 [quant-ph]}\BibitemShut {NoStop}%
\bibitem [{\citenamefont {Lami}\ \emph {et~al.}(2024)\citenamefont {Lami},
  \citenamefont {Haug},\ and\ \citenamefont {Nardis}}]{lami_quantum_2024}%
  \BibitemOpen
  \bibfield  {author} {\bibinfo {author} {\bibfnamefont {G.}~\bibnamefont
  {Lami}}, \bibinfo {author} {\bibfnamefont {T.}~\bibnamefont {Haug}},\ and\
  \bibinfo {author} {\bibfnamefont {J.~D.}\ \bibnamefont {Nardis}},\ }\href
  {http://arxiv.org/abs/2404.18751} {\bibinfo {title} {{Quantum} {State}
  {Designs} with {Clifford} {Enhanced} {Matrix} {Product} {States}}} (\bibinfo
  {year} {2024}),\ \bibinfo {note} {arXiv:2404.18751 [quant-ph]}\BibitemShut
  {NoStop}%
\bibitem [{\citenamefont {Lami}\ and\ \citenamefont
  {Collura}(2024)}]{lami_learning_2024}%
  \BibitemOpen
  \bibfield  {author} {\bibinfo {author} {\bibfnamefont {G.}~\bibnamefont
  {Lami}}\ and\ \bibinfo {author} {\bibfnamefont {M.}~\bibnamefont {Collura}},\
  }\href {http://arxiv.org/abs/2401.16481} {\bibinfo {title} {Learning the
  stabilizer group of a {Matrix} {Product} {State}}} (\bibinfo {year} {2024}),\
  \bibinfo {note} {arXiv:2401.16481 [quant-ph]}\BibitemShut {NoStop}%
\bibitem [{\citenamefont {Mello}\ \emph {et~al.}(2024)\citenamefont {Mello},
  \citenamefont {Santini},\ and\ \citenamefont {Collura}}]{mello_hybrid_2024}%
  \BibitemOpen
  \bibfield  {author} {\bibinfo {author} {\bibfnamefont {A.~F.}\ \bibnamefont
  {Mello}}, \bibinfo {author} {\bibfnamefont {A.}~\bibnamefont {Santini}},\
  and\ \bibinfo {author} {\bibfnamefont {M.}~\bibnamefont {Collura}},\
  }\bibfield  {title} {\bibinfo {title} {Hybrid {Stabilizer} {Matrix} {Product}
  {Operator}},\ }\href {https://doi.org/10.1103/PhysRevLett.133.150604}
  {\bibfield  {journal} {\bibinfo  {journal} {Physical Review Letters}\
  }\textbf {\bibinfo {volume} {133}},\ \bibinfo {pages} {150604} (\bibinfo
  {year} {2024})}\BibitemShut {NoStop}%
\bibitem [{\citenamefont {Qian}\ \emph
  {et~al.}(2024{\natexlab{a}})\citenamefont {Qian}, \citenamefont {Huang},\
  and\ \citenamefont {Qin}}]{qian_clifford_2024}%
  \BibitemOpen
  \bibfield  {author} {\bibinfo {author} {\bibfnamefont {X.}~\bibnamefont
  {Qian}}, \bibinfo {author} {\bibfnamefont {J.}~\bibnamefont {Huang}},\ and\
  \bibinfo {author} {\bibfnamefont {M.}~\bibnamefont {Qin}},\ }\href
  {http://arxiv.org/abs/2407.03202} {\bibinfo {title} {Clifford {Circuits}
  {Augmented} {Time}-{Dependent} {Variational} {Principle}}} (\bibinfo {year}
  {2024}{\natexlab{a}}),\ \bibinfo {note} {arXiv:2407.03202
  [cond-mat]}\BibitemShut {NoStop}%
\bibitem [{\citenamefont {Frau}\ \emph {et~al.}(2024)\citenamefont {Frau},
  \citenamefont {Tarabunga}, \citenamefont {Collura}, \citenamefont
  {Dalmonte},\ and\ \citenamefont {Tirrito}}]{PhysRevB.110.045101}%
  \BibitemOpen
  \bibfield  {author} {\bibinfo {author} {\bibfnamefont {M.}~\bibnamefont
  {Frau}}, \bibinfo {author} {\bibfnamefont {P.~S.}\ \bibnamefont {Tarabunga}},
  \bibinfo {author} {\bibfnamefont {M.}~\bibnamefont {Collura}}, \bibinfo
  {author} {\bibfnamefont {M.}~\bibnamefont {Dalmonte}},\ and\ \bibinfo
  {author} {\bibfnamefont {E.}~\bibnamefont {Tirrito}},\ }\bibfield  {title}
  {\bibinfo {title} {Nonstabilizerness versus entanglement in matrix product
  states},\ }\href {https://doi.org/10.1103/PhysRevB.110.045101} {\bibfield
  {journal} {\bibinfo  {journal} {Phys. Rev. B}\ }\textbf {\bibinfo {volume}
  {110}},\ \bibinfo {pages} {045101} (\bibinfo {year} {2024})}\BibitemShut
  {NoStop}%
\bibitem [{\citenamefont {Tarabunga}\ \emph {et~al.}(2024)\citenamefont
  {Tarabunga}, \citenamefont {Tirrito}, \citenamefont {Ba\~nuls},\ and\
  \citenamefont {Dalmonte}}]{PhysRevLett.133.010601}%
  \BibitemOpen
  \bibfield  {author} {\bibinfo {author} {\bibfnamefont {P.~S.}\ \bibnamefont
  {Tarabunga}}, \bibinfo {author} {\bibfnamefont {E.}~\bibnamefont {Tirrito}},
  \bibinfo {author} {\bibfnamefont {M.~C.}\ \bibnamefont {Ba\~nuls}},\ and\
  \bibinfo {author} {\bibfnamefont {M.}~\bibnamefont {Dalmonte}},\ }\bibfield
  {title} {\bibinfo {title} {Nonstabilizerness via {Matrix} {Product} {States}
  in the {Pauli} {Basis}},\ }\href
  {https://doi.org/10.1103/PhysRevLett.133.010601} {\bibfield  {journal}
  {\bibinfo  {journal} {Phys. Rev. Lett.}\ }\textbf {\bibinfo {volume} {133}},\
  \bibinfo {pages} {010601} (\bibinfo {year} {2024})}\BibitemShut {NoStop}%
\bibitem [{\citenamefont {Dowling}\ \emph {et~al.}(2024)\citenamefont
  {Dowling}, \citenamefont {Kos},\ and\ \citenamefont
  {Turkeshi}}]{dowling2024magic}%
  \BibitemOpen
  \bibfield  {author} {\bibinfo {author} {\bibfnamefont {N.}~\bibnamefont
  {Dowling}}, \bibinfo {author} {\bibfnamefont {P.}~\bibnamefont {Kos}},\ and\
  \bibinfo {author} {\bibfnamefont {X.}~\bibnamefont {Turkeshi}},\ }\bibfield
  {title} {\bibinfo {title} {Magic of the {Heisenberg} {Picture}},\ }\bibfield
  {journal} {\bibinfo  {journal} {arXiv preprint arXiv:2408.16047}\ }\href
  {https://doi.org/https://doi.org/10.48550/arXiv.2408.16047}
  {https://doi.org/10.48550/arXiv.2408.16047} (\bibinfo {year}
  {2024})\BibitemShut {NoStop}%
\bibitem [{\citenamefont {Paviglianiti}\ \emph {et~al.}(2024)\citenamefont
  {Paviglianiti}, \citenamefont {Lami}, \citenamefont {Collura},\ and\
  \citenamefont {Silva}}]{paviglianiti_estimating_2024}%
  \BibitemOpen
  \bibfield  {author} {\bibinfo {author} {\bibfnamefont {A.}~\bibnamefont
  {Paviglianiti}}, \bibinfo {author} {\bibfnamefont {G.}~\bibnamefont {Lami}},
  \bibinfo {author} {\bibfnamefont {M.}~\bibnamefont {Collura}},\ and\ \bibinfo
  {author} {\bibfnamefont {A.}~\bibnamefont {Silva}},\ }\href
  {http://arxiv.org/abs/2405.06054} {\bibinfo {title} {Estimating
  {Non-Stabilizerness} {Dynamics} {Without} {Simulating} {It}}} (\bibinfo
  {year} {2024}),\ \bibinfo {note} {arXiv:2405.06054 [quant-ph]}\BibitemShut
  {NoStop}%
\bibitem [{\citenamefont {Qian}\ \emph
  {et~al.}(2024{\natexlab{b}})\citenamefont {Qian}, \citenamefont {Huang},\
  and\ \citenamefont {Qin}}]{qian_augmenting_2024}%
  \BibitemOpen
  \bibfield  {author} {\bibinfo {author} {\bibfnamefont {X.}~\bibnamefont
  {Qian}}, \bibinfo {author} {\bibfnamefont {J.}~\bibnamefont {Huang}},\ and\
  \bibinfo {author} {\bibfnamefont {M.}~\bibnamefont {Qin}},\ }\href
  {http://arxiv.org/abs/2405.09217} {\bibinfo {title} {Augmenting {Density}
  {Matrix} {Renormalization} {Group} with {Clifford} {Circuits}}} (\bibinfo
  {year} {2024}{\natexlab{b}}),\ \bibinfo {note} {arXiv:2405.09217
  [cond-mat]}\BibitemShut {NoStop}%
\bibitem [{\citenamefont {Huang}\ \emph {et~al.}(2024)\citenamefont {Huang},
  \citenamefont {Qian},\ and\ \citenamefont
  {Qin}}]{huang_non-stabilizerness_2024}%
  \BibitemOpen
  \bibfield  {author} {\bibinfo {author} {\bibfnamefont {J.}~\bibnamefont
  {Huang}}, \bibinfo {author} {\bibfnamefont {X.}~\bibnamefont {Qian}},\ and\
  \bibinfo {author} {\bibfnamefont {M.}~\bibnamefont {Qin}},\ }\href
  {http://arxiv.org/abs/2409.16895} {\bibinfo {title} {Non-stabilizerness
  {Entanglement} {Entropy}: a measure of hardness in the classical simulation
  of quantum many-body systems}} (\bibinfo {year} {2024}),\ \bibinfo {note}
  {arXiv:2409.16895 [quant-ph]}\BibitemShut {NoStop}%
\bibitem [{\citenamefont {Masot-Llima}\ and\ \citenamefont
  {Garcia-Saez}(2024)}]{masot-llima_stabilizer_2024}%
  \BibitemOpen
  \bibfield  {author} {\bibinfo {author} {\bibfnamefont {S.}~\bibnamefont
  {Masot-Llima}}\ and\ \bibinfo {author} {\bibfnamefont {A.}~\bibnamefont
  {Garcia-Saez}},\ }\bibfield  {title} {\bibinfo {title} {Stabilizer tensor
  networks: Universal quantum simulator on a basis of stabilizer states},\
  }\href@noop {} {\bibfield  {journal} {\bibinfo  {journal} {Phys. Rev. Lett.}\
  }\textbf {\bibinfo {volume} {133}},\ \bibinfo {pages} {230601} (\bibinfo
  {year} {2024})}\BibitemShut {NoStop}%
\bibitem [{\citenamefont {Fowler}\ and\ \citenamefont
  {Gidney}(2019)}]{fowler_low_2019}%
  \BibitemOpen
  \bibfield  {author} {\bibinfo {author} {\bibfnamefont {A.~G.}\ \bibnamefont
  {Fowler}}\ and\ \bibinfo {author} {\bibfnamefont {C.}~\bibnamefont
  {Gidney}},\ }\href {http://arxiv.org/abs/1808.06709} {\bibinfo {title} {Low
  overhead quantum computation using lattice surgery}} (\bibinfo {year}
  {2019}),\ \bibinfo {note} {arXiv:1808.06709 [quant-ph]}\BibitemShut {NoStop}%
\bibitem [{\citenamefont {Reardon-Smith}\ \emph {et~al.}(2024)\citenamefont
  {Reardon-Smith}, \citenamefont {Oszmaniec},\ and\ \citenamefont
  {Korzekwa}}]{Reardon_Smith_2024}%
  \BibitemOpen
  \bibfield  {author} {\bibinfo {author} {\bibfnamefont {O.}~\bibnamefont
  {Reardon-Smith}}, \bibinfo {author} {\bibfnamefont {M.}~\bibnamefont
  {Oszmaniec}},\ and\ \bibinfo {author} {\bibfnamefont {K.}~\bibnamefont
  {Korzekwa}},\ }\bibfield  {title} {\bibinfo {title} {Improved simulation of
  quantum circuits dominated by free fermionic operations},\ }\href
  {https://doi.org/10.22331/q-2024-12-04-1549} {\bibfield  {journal} {\bibinfo
  {journal} {Quantum}\ }\textbf {\bibinfo {volume} {8}},\ \bibinfo {pages}
  {1549} (\bibinfo {year} {2024})}\BibitemShut {NoStop}%
\bibitem [{\citenamefont {True}\ and\ \citenamefont
  {Hamma}(2022)}]{true_transitions_2022}%
  \BibitemOpen
  \bibfield  {author} {\bibinfo {author} {\bibfnamefont {S.}~\bibnamefont
  {True}}\ and\ \bibinfo {author} {\bibfnamefont {A.}~\bibnamefont {Hamma}},\
  }\bibfield  {title} {\bibinfo {title} {Transitions in {Entanglement}
  {Complexity} in {Random} {Circuits}},\ }\href
  {https://doi.org/10.22331/q-2022-09-22-818} {\bibfield  {journal} {\bibinfo
  {journal} {Quantum}\ }\textbf {\bibinfo {volume} {6}},\ \bibinfo {pages}
  {818} (\bibinfo {year} {2022})}\BibitemShut {NoStop}%
\bibitem [{\citenamefont {Leone}\ \emph
  {et~al.}(2024{\natexlab{a}})\citenamefont {Leone}, \citenamefont {Oliviero},\
  and\ \citenamefont {Hamma}}]{leone_learning_2024}%
  \BibitemOpen
  \bibfield  {author} {\bibinfo {author} {\bibfnamefont {L.}~\bibnamefont
  {Leone}}, \bibinfo {author} {\bibfnamefont {S.~F.~E.}\ \bibnamefont
  {Oliviero}},\ and\ \bibinfo {author} {\bibfnamefont {A.}~\bibnamefont
  {Hamma}},\ }\bibfield  {title} {\bibinfo {title} {Learning t-doped stabilizer
  states},\ }\href {https://doi.org/10.22331/q-2024-05-27-1361} {\bibfield
  {journal} {\bibinfo  {journal} {Quantum}\ }\textbf {\bibinfo {volume} {8}},\
  \bibinfo {pages} {1361} (\bibinfo {year} {2024}{\natexlab{a}})},\ \bibinfo
  {note} {arXiv:2305.15398 [quant-ph]}\BibitemShut {NoStop}%
\bibitem [{\citenamefont {Roetteler}(2009)}]{roetteler_quantum_2009}%
  \BibitemOpen
  \bibfield  {author} {\bibinfo {author} {\bibfnamefont {M.}~\bibnamefont
  {Roetteler}},\ }\href {http://arxiv.org/abs/0811.3208} {\bibinfo {title}
  {Quantum algorithms for highly non-linear {Boolean} functions}} (\bibinfo
  {year} {2009}),\ \bibinfo {note} {arXiv:0811.3208 [quant-ph]}\BibitemShut
  {NoStop}%
\bibitem [{\citenamefont {Kissinger}\ \emph {et~al.}(2022)\citenamefont
  {Kissinger}, \citenamefont {Wetering},\ and\ \citenamefont
  {Vilmart}}]{kissinger_classical_2022}%
  \BibitemOpen
  \bibfield  {author} {\bibinfo {author} {\bibfnamefont {A.}~\bibnamefont
  {Kissinger}}, \bibinfo {author} {\bibfnamefont {J.~v.~d.}\ \bibnamefont
  {Wetering}},\ and\ \bibinfo {author} {\bibfnamefont {R.}~\bibnamefont
  {Vilmart}},\ }\bibfield  {title} {\bibinfo {title} {Classical simulation of
  quantum circuits with partial and graphical stabiliser decompositions},\
  }\href {https://doi.org/10.4230/LIPIcs.TQC.2022.5} {\bibfield  {journal}
  {\bibinfo  {journal} {LIPIcs, Volume 232, TQC 2022}\ }\textbf {\bibinfo
  {volume} {232}},\ \bibinfo {pages} {5:1} (\bibinfo {year} {2022})},\ \bibinfo
  {note} {arXiv:2202.09202 [quant-ph]}\BibitemShut {NoStop}%
\bibitem [{\citenamefont {Kissinger}\ and\ \citenamefont {Van
  De~Wetering}(2022)}]{kissinger_simulating_2022}%
  \BibitemOpen
  \bibfield  {author} {\bibinfo {author} {\bibfnamefont {A.}~\bibnamefont
  {Kissinger}}\ and\ \bibinfo {author} {\bibfnamefont {J.}~\bibnamefont {Van
  De~Wetering}},\ }\bibfield  {title} {\bibinfo {title} {Simulating quantum
  circuits with {ZX}-calculus reduced stabiliser decompositions},\ }\href
  {https://doi.org/10.1088/2058-9565/ac5d20} {\bibfield  {journal} {\bibinfo
  {journal} {Quantum Science and Technology}\ }\textbf {\bibinfo {volume}
  {7}},\ \bibinfo {pages} {044001} (\bibinfo {year} {2022})}\BibitemShut
  {NoStop}%
\bibitem [{\citenamefont {Aaronson}\ and\ \citenamefont
  {Gottesman}(2008)}]{aaronson_improved_2008}%
  \BibitemOpen
  \bibfield  {author} {\bibinfo {author} {\bibfnamefont {S.}~\bibnamefont
  {Aaronson}}\ and\ \bibinfo {author} {\bibfnamefont {D.}~\bibnamefont
  {Gottesman}},\ }\href {https://doi.org/10.48550/arXiv.quant-ph/0406196}
  {\bibinfo {title} {Improved {Simulation} of {Stabilizer} {Circuits}}}
  (\bibinfo {year} {2008}),\ \bibinfo {note}
  {arXiv:quant-ph/0406196}\BibitemShut {NoStop}%
\bibitem [{\citenamefont {Dupont}\ \emph {et~al.}(2022)\citenamefont {Dupont},
  \citenamefont {Didier}, \citenamefont {Hodson}, \citenamefont {Moore},\ and\
  \citenamefont {Reagor}}]{dupont_entanglement_2022}%
  \BibitemOpen
  \bibfield  {author} {\bibinfo {author} {\bibfnamefont {M.}~\bibnamefont
  {Dupont}}, \bibinfo {author} {\bibfnamefont {N.}~\bibnamefont {Didier}},
  \bibinfo {author} {\bibfnamefont {M.~J.}\ \bibnamefont {Hodson}}, \bibinfo
  {author} {\bibfnamefont {J.~E.}\ \bibnamefont {Moore}},\ and\ \bibinfo
  {author} {\bibfnamefont {M.~J.}\ \bibnamefont {Reagor}},\ }\bibfield  {title}
  {\bibinfo {title} {Entanglement perspective on the quantum approximate
  optimization algorithm},\ }\href
  {https://doi.org/10.1103/PhysRevA.106.022423} {\bibfield  {journal} {\bibinfo
   {journal} {Physical Review A}\ }\textbf {\bibinfo {volume} {106}},\ \bibinfo
  {pages} {022423} (\bibinfo {year} {2022})}\BibitemShut {NoStop}%
\bibitem [{\citenamefont {{Qiskit contributors}}(2023)}]{qiskit}%
  \BibitemOpen
  \bibfield  {author} {\bibinfo {author} {\bibnamefont {{Qiskit
  contributors}}},\ }\href {https://doi.org/10.5281/zenodo.2573505} {\bibinfo
  {title} {Qiskit: {An} {Open}-source {Framework} for {Quantum} {Computing}}}
  (\bibinfo {year} {2023})\BibitemShut {NoStop}%
\bibitem [{\citenamefont {Bravyi}\ and\ \citenamefont
  {Maslov}(2021)}]{bravyi2021hadamard}%
  \BibitemOpen
  \bibfield  {author} {\bibinfo {author} {\bibfnamefont {S.}~\bibnamefont
  {Bravyi}}\ and\ \bibinfo {author} {\bibfnamefont {D.}~\bibnamefont
  {Maslov}},\ }\bibfield  {title} {\bibinfo {title} {Hadamard-free circuits
  expose the structure of the {Clifford} group},\ }\href
  {https://doi.org/10.1109/TIT.2021.3081415} {\bibfield  {journal} {\bibinfo
  {journal} {IEEE Transactions on Information Theory}\ }\textbf {\bibinfo
  {volume} {67}},\ \bibinfo {pages} {4546} (\bibinfo {year}
  {2021})}\BibitemShut {NoStop}%
\bibitem [{\citenamefont {Aaronson}\ and\ \citenamefont
  {Gottesman}(2004)}]{aaronson2004improved}%
  \BibitemOpen
  \bibfield  {author} {\bibinfo {author} {\bibfnamefont {S.}~\bibnamefont
  {Aaronson}}\ and\ \bibinfo {author} {\bibfnamefont {D.}~\bibnamefont
  {Gottesman}},\ }\bibfield  {title} {\bibinfo {title} {Improved simulation of
  stabilizer circuits},\ }\href {https://doi.org/10.1103/PhysRevA.70.052328}
  {\bibfield  {journal} {\bibinfo  {journal} {Physical Review A}\ }\textbf
  {\bibinfo {volume} {70}},\ \bibinfo {pages} {052328} (\bibinfo {year}
  {2004})}\BibitemShut {NoStop}%
\bibitem [{\citenamefont {Eisert}(2013)}]{eisert_entanglement_2013}%
  \BibitemOpen
  \bibfield  {author} {\bibinfo {author} {\bibfnamefont {J.}~\bibnamefont
  {Eisert}},\ }\href {http://arxiv.org/abs/1308.3318} {\bibinfo {title}
  {Entanglement and tensor network states}} (\bibinfo {year} {2013}),\ \bibinfo
  {note} {arXiv:1308.3318 [quant-ph]}\BibitemShut {NoStop}%
\bibitem [{\citenamefont {Bremner}\ \emph {et~al.}(2011)\citenamefont
  {Bremner}, \citenamefont {Jozsa},\ and\ \citenamefont
  {Shepherd}}]{bremner2011classical}%
  \BibitemOpen
  \bibfield  {author} {\bibinfo {author} {\bibfnamefont {M.~J.}\ \bibnamefont
  {Bremner}}, \bibinfo {author} {\bibfnamefont {R.}~\bibnamefont {Jozsa}},\
  and\ \bibinfo {author} {\bibfnamefont {D.~J.}\ \bibnamefont {Shepherd}},\
  }\bibfield  {title} {\bibinfo {title} {Classical simulation of commuting
  quantum computations implies collapse of the polynomial hierarchy},\ }\href
  {https://royalsocietypublishing.org/doi/abs/10.1098/rspa.2010.0301}
  {\bibfield  {journal} {\bibinfo  {journal} {Proceedings of the Royal Society
  A: Mathematical, Physical and Engineering Sciences}\ }\textbf {\bibinfo
  {volume} {467}},\ \bibinfo {pages} {459} (\bibinfo {year}
  {2011})}\BibitemShut {NoStop}%
\bibitem [{\citenamefont {Leone}\ \emph {et~al.}(2021)\citenamefont {Leone},
  \citenamefont {Oliviero}, \citenamefont {Zhou},\ and\ \citenamefont
  {Hamma}}]{Leone2021quantumchaosis}%
  \BibitemOpen
  \bibfield  {author} {\bibinfo {author} {\bibfnamefont {L.}~\bibnamefont
  {Leone}}, \bibinfo {author} {\bibfnamefont {S.~F.~E.}\ \bibnamefont
  {Oliviero}}, \bibinfo {author} {\bibfnamefont {Y.}~\bibnamefont {Zhou}},\
  and\ \bibinfo {author} {\bibfnamefont {A.}~\bibnamefont {Hamma}},\ }\bibfield
   {title} {\bibinfo {title} {Quantum {C}haos is {Q}uantum},\ }\href
  {https://doi.org/10.22331/q-2021-05-04-453} {\bibfield  {journal} {\bibinfo
  {journal} {{Quantum}}\ }\textbf {\bibinfo {volume} {5}},\ \bibinfo {pages}
  {453} (\bibinfo {year} {2021})}\BibitemShut {NoStop}%
\bibitem [{\citenamefont {Leone}\ \emph
  {et~al.}(2024{\natexlab{b}})\citenamefont {Leone}, \citenamefont {Oliviero},
  \citenamefont {Lloyd},\ and\ \citenamefont {Hamma}}]{PhysRevA.109.022429}%
  \BibitemOpen
  \bibfield  {author} {\bibinfo {author} {\bibfnamefont {L.}~\bibnamefont
  {Leone}}, \bibinfo {author} {\bibfnamefont {S.~F.~E.}\ \bibnamefont
  {Oliviero}}, \bibinfo {author} {\bibfnamefont {S.}~\bibnamefont {Lloyd}},\
  and\ \bibinfo {author} {\bibfnamefont {A.}~\bibnamefont {Hamma}},\ }\bibfield
   {title} {\bibinfo {title} {Learning efficient decoders for quasichaotic
  quantum scramblers},\ }\href {https://doi.org/10.1103/PhysRevA.109.022429}
  {\bibfield  {journal} {\bibinfo  {journal} {Phys. Rev. A}\ }\textbf {\bibinfo
  {volume} {109}},\ \bibinfo {pages} {022429} (\bibinfo {year}
  {2024}{\natexlab{b}})}\BibitemShut {NoStop}%
\bibitem [{\citenamefont {Oliviero}\ \emph {et~al.}(2024)\citenamefont
  {Oliviero}, \citenamefont {Leone}, \citenamefont {Lloyd},\ and\ \citenamefont
  {Hamma}}]{PhysRevLett.132.080402}%
  \BibitemOpen
  \bibfield  {author} {\bibinfo {author} {\bibfnamefont {S.~F.~E.}\
  \bibnamefont {Oliviero}}, \bibinfo {author} {\bibfnamefont {L.}~\bibnamefont
  {Leone}}, \bibinfo {author} {\bibfnamefont {S.}~\bibnamefont {Lloyd}},\ and\
  \bibinfo {author} {\bibfnamefont {A.}~\bibnamefont {Hamma}},\ }\bibfield
  {title} {\bibinfo {title} {Unscrambling quantum information with clifford
  decoders},\ }\href {https://doi.org/10.1103/PhysRevLett.132.080402}
  {\bibfield  {journal} {\bibinfo  {journal} {Phys. Rev. Lett.}\ }\textbf
  {\bibinfo {volume} {132}},\ \bibinfo {pages} {080402} (\bibinfo {year}
  {2024})}\BibitemShut {NoStop}%
\bibitem [{\citenamefont {Gu}\ \emph {et~al.}(2024)\citenamefont {Gu},
  \citenamefont {Oliviero},\ and\ \citenamefont
  {Leone}}]{gu2024magicinducedcomp}%
  \BibitemOpen
  \bibfield  {author} {\bibinfo {author} {\bibfnamefont {A.}~\bibnamefont
  {Gu}}, \bibinfo {author} {\bibfnamefont {S.~F.~E.}\ \bibnamefont
  {Oliviero}},\ and\ \bibinfo {author} {\bibfnamefont {L.}~\bibnamefont
  {Leone}},\ }\href@noop {} {\bibinfo {title} {Magic-induced computational
  separation in entanglement theory}} (\bibinfo {year} {2024}),\ \Eprint
  {https://arxiv.org/abs/2403.19610} {arXiv:2403.19610 [quant-ph]} \BibitemShut
  {NoStop}%
\bibitem [{\citenamefont {Mele}\ \emph {et~al.}(2024)\citenamefont {Mele},
  \citenamefont {Angrisani}, \citenamefont {Ghosh}, \citenamefont {Khatri},
  \citenamefont {Eisert}, \citenamefont {Fran{\c{c}}a},\ and\ \citenamefont
  {Quek}}]{mele2024noise}%
  \BibitemOpen
  \bibfield  {author} {\bibinfo {author} {\bibfnamefont {A.~A.}\ \bibnamefont
  {Mele}}, \bibinfo {author} {\bibfnamefont {A.}~\bibnamefont {Angrisani}},
  \bibinfo {author} {\bibfnamefont {S.}~\bibnamefont {Ghosh}}, \bibinfo
  {author} {\bibfnamefont {S.}~\bibnamefont {Khatri}}, \bibinfo {author}
  {\bibfnamefont {J.}~\bibnamefont {Eisert}}, \bibinfo {author} {\bibfnamefont
  {D.~S.}\ \bibnamefont {Fran{\c{c}}a}},\ and\ \bibinfo {author} {\bibfnamefont
  {Y.}~\bibnamefont {Quek}},\ }\bibfield  {title} {\bibinfo {title}
  {Noise-induced shallow circuits and absence of barren plateaus},\ }\href
  {https://arxiv.org/abs/2403.13927} {\bibfield  {journal} {\bibinfo  {journal}
  {arXiv preprint arXiv:2403.13927}\ } (\bibinfo {year} {2024})}\BibitemShut
  {NoStop}%
\bibitem [{\citenamefont {Amy}\ and\ \citenamefont
  {Stinchcombe}(2024)}]{amy_polynomial-time_2024}%
  \BibitemOpen
  \bibfield  {author} {\bibinfo {author} {\bibfnamefont {M.}~\bibnamefont
  {Amy}}\ and\ \bibinfo {author} {\bibfnamefont {L.~S.}\ \bibnamefont
  {Stinchcombe}},\ }\href {http://arxiv.org/abs/2408.02778} {\bibinfo {title}
  {{Polynomial}-{Time} {Classical} {Simulation} of {Hidden} {Shift} {Circuits}
  via {Confluent} {Rewriting} of {Symbolic} {Sums}}} (\bibinfo {year} {2024}),\
  \bibinfo {note} {arXiv:2408.02778 [quant-ph]}\BibitemShut {NoStop}%
\bibitem [{\citenamefont {Fux}\ \emph {et~al.}(2024)\citenamefont {Fux},
  \citenamefont {Béri}, \citenamefont {Fazio},\ and\ \citenamefont
  {Tirrito}}]{fux_disentangling_2024}%
  \BibitemOpen
  \bibfield  {author} {\bibinfo {author} {\bibfnamefont {G.~E.}\ \bibnamefont
  {Fux}}, \bibinfo {author} {\bibfnamefont {B.}~\bibnamefont {Béri}}, \bibinfo
  {author} {\bibfnamefont {R.}~\bibnamefont {Fazio}},\ and\ \bibinfo {author}
  {\bibfnamefont {E.}~\bibnamefont {Tirrito}},\ }\href
  {http://arxiv.org/abs/2410.09001} {\bibinfo {title} {Disentangling unitary
  dynamics with classically simulable quantum circuits}} (\bibinfo {year}
  {2024}),\ \bibinfo {note} {arXiv:2410.09001 [quant-ph]}\BibitemShut {NoStop}%
\bibitem [{\citenamefont {Preskill}(2018)}]{preskill2018quantum}%
  \BibitemOpen
  \bibfield  {author} {\bibinfo {author} {\bibfnamefont {J.}~\bibnamefont
  {Preskill}},\ }\bibfield  {title} {\bibinfo {title} {Quantum {Computing} in
  the {NISQ} era and beyond},\ }\href
  {https://doi.org/10.22331/q-2018-08-06-79} {\bibfield  {journal} {\bibinfo
  {journal} {Quantum}\ }\textbf {\bibinfo {volume} {2}},\ \bibinfo {pages} {79}
  (\bibinfo {year} {2018})}\BibitemShut {NoStop}%
\bibitem [{\citenamefont {Farhi}\ \emph {et~al.}(2014)\citenamefont {Farhi},
  \citenamefont {Goldstone},\ and\ \citenamefont {Gutmann}}]{farhi2014quantum}%
  \BibitemOpen
  \bibfield  {author} {\bibinfo {author} {\bibfnamefont {E.}~\bibnamefont
  {Farhi}}, \bibinfo {author} {\bibfnamefont {J.}~\bibnamefont {Goldstone}},\
  and\ \bibinfo {author} {\bibfnamefont {S.}~\bibnamefont {Gutmann}},\
  }\bibfield  {title} {\bibinfo {title} {A {Quantum} {Approximate}
  {Optimization} {Algorithm}},\ }\href {https://arxiv.org/abs/1411.4028}
  {\bibfield  {journal} {\bibinfo  {journal} {arXiv preprint arXiv:1411.4028}\
  } (\bibinfo {year} {2014})}\BibitemShut {NoStop}%
\bibitem [{\citenamefont {Peruzzo}\ \emph {et~al.}(2014)\citenamefont
  {Peruzzo}, \citenamefont {McClean}, \citenamefont {Shadbolt}, \citenamefont
  {Yung}, \citenamefont {Zhou}, \citenamefont {Love}, \citenamefont
  {Aspuru-Guzik},\ and\ \citenamefont {O’Brien}}]{peruzzo2014variational}%
  \BibitemOpen
  \bibfield  {author} {\bibinfo {author} {\bibfnamefont {A.}~\bibnamefont
  {Peruzzo}}, \bibinfo {author} {\bibfnamefont {J.}~\bibnamefont {McClean}},
  \bibinfo {author} {\bibfnamefont {P.}~\bibnamefont {Shadbolt}}, \bibinfo
  {author} {\bibfnamefont {M.-H.}\ \bibnamefont {Yung}}, \bibinfo {author}
  {\bibfnamefont {X.-Q.}\ \bibnamefont {Zhou}}, \bibinfo {author}
  {\bibfnamefont {P.~J.}\ \bibnamefont {Love}}, \bibinfo {author}
  {\bibfnamefont {A.}~\bibnamefont {Aspuru-Guzik}},\ and\ \bibinfo {author}
  {\bibfnamefont {J.~L.}\ \bibnamefont {O’Brien}},\ }\bibfield  {title}
  {\bibinfo {title} {A variational eigenvalue solver on a photonic quantum
  processor},\ }\href {https://doi.org/10.1038/ncomms5213} {\bibfield
  {journal} {\bibinfo  {journal} {Nature {C}ommunications}\ }\textbf {\bibinfo
  {volume} {5}},\ \bibinfo {pages} {1} (\bibinfo {year} {2014})}\BibitemShut
  {NoStop}%
\bibitem [{\citenamefont {Qassim}\ \emph {et~al.}(2021)\citenamefont {Qassim},
  \citenamefont {Pashayan},\ and\ \citenamefont
  {Gosset}}]{Qassim2021improvedupperbounds}%
  \BibitemOpen
  \bibfield  {author} {\bibinfo {author} {\bibfnamefont {H.}~\bibnamefont
  {Qassim}}, \bibinfo {author} {\bibfnamefont {H.}~\bibnamefont {Pashayan}},\
  and\ \bibinfo {author} {\bibfnamefont {D.}~\bibnamefont {Gosset}},\
  }\bibfield  {title} {\bibinfo {title} {Improved upper bounds on the
  stabilizer rank of magic states},\ }\href
  {https://doi.org/10.22331/q-2021-12-20-606} {\bibfield  {journal} {\bibinfo
  {journal} {{Quantum}}\ }\textbf {\bibinfo {volume} {5}},\ \bibinfo {pages}
  {606} (\bibinfo {year} {2021})}\BibitemShut {NoStop}%
\bibitem [{\citenamefont {{Harper, B, Nakhl, Azar C.}}(2025)}]{mastgit}%
  \BibitemOpen
  \bibfield  {author} {\bibinfo {author} {\bibnamefont {{Harper, B, Nakhl, Azar
  C.}}},\ }\href@noop {} {\bibinfo {title} {Magic {State} {Injection}
  {Augmented} {Stabilizer} {Tensor} {Networks}}},\ \bibinfo {howpublished}
  {\url{https://github.com/aristaeus/mast}} (\bibinfo {year}
  {2025})\BibitemShut {NoStop}%
\bibitem [{\citenamefont {Gottesman}(1997)}]{gottesman1997stabilizer}%
  \BibitemOpen
  \bibfield  {author} {\bibinfo {author} {\bibfnamefont {D.}~\bibnamefont
  {Gottesman}},\ }\href@noop {} {\emph {\bibinfo {title} {Stabilizer {Codes}
  and {Quantum} {Error} {Correction}}}}\ (\bibinfo  {publisher} {California
  Institute of Technology},\ \bibinfo {year} {1997})\BibitemShut {NoStop}%
\bibitem [{\citenamefont {Or{\'u}s}(2014)}]{orus2014practical}%
  \BibitemOpen
  \bibfield  {author} {\bibinfo {author} {\bibfnamefont {R.}~\bibnamefont
  {Or{\'u}s}},\ }\bibfield  {title} {\bibinfo {title} {A {Practical}
  {Introduction} to {Tensor} {Networks}: {Matrix} {Product} {States} and
  {Projected} {Entangled} {Pair} {States}},\ }\href
  {https://doi.org/10.1016/j.aop.2014.06.013} {\bibfield  {journal} {\bibinfo
  {journal} {Annals of Physics}\ }\textbf {\bibinfo {volume} {349}},\ \bibinfo
  {pages} {117} (\bibinfo {year} {2014})}\BibitemShut {NoStop}%
\bibitem [{\citenamefont {Nielsen}\ and\ \citenamefont
  {Chuang}(2000)}]{nielsen2000quantum}%
  \BibitemOpen
  \bibfield  {author} {\bibinfo {author} {\bibfnamefont {M.~A.}\ \bibnamefont
  {Nielsen}}\ and\ \bibinfo {author} {\bibfnamefont {I.~L.}\ \bibnamefont
  {Chuang}},\ }\href@noop {} {\emph {\bibinfo {title} {Quantum {Computation}
  and {Quantum} {Information}}}}\ (\bibinfo  {publisher} {Cambridge University
  Press},\ \bibinfo {address} {Cambridge},\ \bibinfo {year} {2000})\BibitemShut
  {NoStop}%
\bibitem [{\citenamefont {Koenig}\ and\ \citenamefont
  {Smolin}(2014)}]{10.1063/1.4903507}%
  \BibitemOpen
  \bibfield  {author} {\bibinfo {author} {\bibfnamefont {R.}~\bibnamefont
  {Koenig}}\ and\ \bibinfo {author} {\bibfnamefont {J.~A.}\ \bibnamefont
  {Smolin}},\ }\bibfield  {title} {\bibinfo {title} {How to efficiently select
  an arbitrary clifford group element},\ }\href
  {https://doi.org/10.1063/1.4903507} {\bibfield  {journal} {\bibinfo
  {journal} {Journal of Mathematical Physics}\ }\textbf {\bibinfo {volume}
  {55}},\ \bibinfo {pages} {122202} (\bibinfo {year} {2014})},\ \Eprint
  {https://arxiv.org/abs/https://pubs.aip.org/aip/jmp/article-pdf/doi/10.1063/1.4903507/15975983/122202\_1\_online.pdf}
  {https://pubs.aip.org/aip/jmp/article-pdf/doi/10.1063/1.4903507/15975983/122202\_1\_online.pdf}
  \BibitemShut {NoStop}%
\end{thebibliography}%

\appendix
\section{Stabilizer state and Matrix Product State Background} \label{appendix:prelim}
\subsection{Stabilizer states}
Stabilizer states $\ket{\phi}$ are distinguished states that can be completely characterized by specifying an associated stabilizer group $\mathcal{S}$~\cite{gottesman1997stabilizer}. More precisely,
the stabilizer group $\mathcal{S}$ is defined to be the maximal subgroup of the Pauli group $\mathcal{P}$ over $N$ qubits such that $S\ket{\phi} = \ket{\phi}$ for all $S\in \mathcal{S}$. From a general theoretical perspective, stabilizers play an important role in error correction as they define a set of operators that are invariant under logical operations~\cite{gottesman1997stabilizer}.

More relevant to our discussion however, is their role in the simulation of circuits consisting only of Clifford gates. Stabilizer states being uniquely defined by their stabilizer group, they can be encoded efficiently in an $N \times N$ matrix called the stabilizer tableau, where each row is a generator for the stabilizer group. When applying a Clifford gate $C$ to a stabilizer state, we can find a new set of stabilizers for the updated state by conjugating the stabilizer group $\mathcal{S}$ by the Clifford operation,
\begin{align*}
    C \ket{\phi} &= C S \ket{\phi} \\
                 &= C S C^{-1} \left(C \ket{\phi}\right).
\end{align*}
When $C$ is a Clifford gate, conjugating $S\in\mathcal{S}$ in this way results in a Pauli string~\cite{gottesman1998heisenberg}, which is the updated stabilizer.

Related to the stabilizer group is the destabilizer group $\mathcal{D} = \mathcal{P} /\ \mathcal{S}$, that is the remaining Paulis needed to form the whole group. Although not essential for stabilizer simulators, keeping track of the remaining generators for the Pauli group allows more efficient simulation of projective measurement in a stabilizer simulator~\cite{aaronson2004improved}.

\subsection{Matrix Product States}
Matrix Product States (MPS) are a tensor network representation of one-dimensional many body quantum systems. For a 2-level system (qubits) whose Hilbert space is spanned by the basis states $\ket{s}$ where $s$ is a binary string representation of some number $1,\dots,2^N$, the MPS may be represented as
\begin{equation}
    \ket{\psi} = \sum_{s_1,s_2,\ldots,s_N} A_{s_1}^{(1)} A_{s_2}^{(2)} \dots A_{s_N}^{(N)}\ket{s_1s_2\dots s_N} \label{eq:mps}
\end{equation}
where $A_{s_i}^{(i)}$ are matrices which can be optimally chosen to be of size $\chi_{i-1} \times \chi_{i}$,
with the matrices at the end being of size $1 \times \chi_1$ and $\chi_{N} \times 1$ respectively,
where $\chi_i$ is referred to as the bond-dimension (between sites $i$ and $i+1$) and describes the bipartite entanglement of the state at that site.
One may find a lower-entanglement approximation to the state $\ket{\psi}$ by performing a Singular Value Decomposition (SVD) on the $A_{s_i}^{(i)}$, keeping only a fixed number of singular values.
A more detailed introduction to MPS may be found in~\cite{schollwock2011density}.

For simplicity for the upcoming section, we will set $\nu_i = \text{Tr}(A_{s_1}^{(1)} A_{s_2}^{(2)}\cdots A_{s_N}^{(N)})$ where $i$ is the decimal representation of the binary string $s_1s_2\cdots s_N$. Hence Equation \eqref{eq:mps} can be written as
\begin{equation}
    \ket{\psi} = \sum_{i = 1}^{2^N} \nu_i \ket{i}.
\end{equation}

\section{Stabilizer Tensor Networks} \label{appendix:stn}
Following the work in~\cite{masot-llima_stabilizer_2024} that established the Stabilizer Tensor Network (STN) protocol, in this section we describe the foundational aspects of this simulation protocol. Namely we describe how this can be used to describe a general quantum state, and how it can be used to perform both Clifford and non-Clifford operations, as well as expectation value calculations and projective measurements. 

A quantum state $\ket{\psi}$ in this protocol is represented in the following form,
\begin{equation}
    \ket{\psi} = \sum_{i=1}^{2^N} \nu_i D_{\hat{\imath}} \ket{\phi} \label{eq:stn}
\end{equation}
where $\ket{\phi}$ is a stabilizer state stored as a tableau and updated as discussed below, and $\hat{\imath}=(s_1,s_2\dots s_N)$ as above. The operator $D_{\hat{\imath}}$ is created from the vector $\hat{\imath}$,
\begin{equation}
    D_{\hat{\imath}} = \prod_{j = 1}^N \phi_j^{\hat{\imath}_j} \label{eq:destab_op}
\end{equation}
where $\phi_j$ is the Pauli string in the $j$\textsuperscript{th} row of $\ket{\phi}$'s stabilizer tableau. As $\mathcal{S} \cup \mathcal{D}$ spans ${\rm End}((\mbc^2)^{\otimes N}) $, any $N$ qubit state can be representated in the form of Equation~\eqref{eq:stn}.

\subsection{Clifford Operations}
To apply a Clifford operation $C$ to the state $\ket{\psi}$, we conjugate the operator $D_{\hat{\imath}}$ by $C$
\begin{align*}
    C\ket{\psi} = C\sum_i \nu_i D_{\hat{\imath}} \ket{\phi} &= \sum_i \nu_i C C^{-1} \tilde{D}_{\hat{\imath}} C \ket{\phi} \\
                &= \sum_i \nu_i \tilde{D}_{\hat{\imath}} \tilde{\ket{\phi}}
\end{align*}
where $\tilde{D}_{\hat{\imath}} = C D_{\hat{\imath}} C^{-1}$. Recall that Clifford conjugation is the update rule for a stabilizer tableau simulator~\cite{gottesman1998heisenberg}, so updating the tableau of $\ket{\phi}$ is sufficient to update $D_{\hat{\imath}}$ as well. Hence Clifford operations only update $\ket{\phi}$, using the standard stabilizer update rules~\cite{aaronson2004improved}, and leave the MPS unchanged.

\subsection{Non-Clifford Operations} \label{appendix:stn_non}
To perform a non-Clifford operation $U$, first one must find a decomposition of the following form
\begin{equation}
    U = \sum_i c_i D_{\hat{d}_i} S_{\hat{s}_i} \label{eq:udecomp}
\end{equation}
where $c_i$ are complex coefficients, and $\hat{d}_i$ and $\hat{s}_i$ are boolean vectors that like $\hat{\imath}$ pick out (de)stabilizer rows. These can be found by starting with a decomposition of $U$ in terms of Pauli strings $P_i$,
\begin{equation*}
    U = \sum_i a_i P_i, \quad \text{where } P_i = D_{\hat{d}_i} S_{\hat{s}_i}
\end{equation*}
Each $P_i$ must then be decomposed into products of rows in the (de)stabilizer tableau. We proceed from Proposition 3 in~\cite{aaronson2004improved}, which shows that an element ($D$) $S$ in the (de)stabilizer may be written as a product of the (de)stabilizer terms that correspond to rows in the (stabilizer) destabilizer that anticommute with the operator ($D$) $S$. Thus, to find $D_{\hat{d}_i}$ and $S_{\hat{s}_i}$ from Equation (\ref{eq:udecomp}), we simply check whether $P_i$ anticommutes with each row in the (de)stabilizer tableau, select the corresponding tableau row, and perform some bookkeeping to keep track of the phase.

After finding the decomposition, one may apply the operator to the STN as follows
\begin{align}
    U\ket{\psi} &= \sum_j c_j D_{d_{\hat{\jmath}}} S_{s_{\hat{\jmath}}} \sum_{i=1}^{2^N} \nu_i D_{\hat{\imath}} \ket{\phi} \\
    &= \sum_j \sum_{i=1}^{2^N} c_j \nu_i D_{d_{\hat{\jmath}}} S_{s_{\hat{\jmath}}} D_{\hat{\imath}} \ket{\phi} \\
    &= \sum_j \sum_{i=1}^{2^N} c_j \nu_i (-1)^{s_{\hat{\jmath}} \cdot \hat{\imath} } D_{d_{\hat{\jmath}} +\hat{\imath}} \ket{\phi} 
\end{align}
where we note that the $i$\textsuperscript{th} stabilizer and $i$\textsuperscript{th} destabilizer anti-commute by definition. Also note that $D_{\hat{\jmath}+\hat{\imath}} = D_{\hat{\jmath}}D_{\hat{\imath}}$ for any $\hat{\imath}$ and $\hat{\jmath}$ directly from Equation \eqref{eq:destab_op}. We note, that unlike Clifford operations that only update the stabilizer state $\ket{\phi}$, non-Clifford operations appear to leave the stabilizer state unchanged, and result instead in an update to the MPS. 

Continuing to follow~\cite{masot-llima_stabilizer_2024}, to see how the MPS is updated, consider a decomposition with only two terms $U=c_1 D_{\hat{d}_1}S_{\hat{s}_1} + c_2D_{\hat{d}_2}S_{\hat{s}_2}$. We can write this as
\begin{equation}
    U = (c_1 I + c_2 D_{\hat{d}_2}S_{\hat{s}_2} S_{\hat{s}_1}D_{\hat{d}_1}) D_{\hat{d}_1}S_{\hat{s}_1}
\end{equation}
recognising that $S_{\hat{\imath}} S_{\hat{\imath}} = D_{\hat{\imath}} D_{\hat{\imath}} = I$ for any boolean vector $\hat{\imath}$. We note that the $D_{\hat{d}_1}S_{\hat{s}_1}$ on the right hand side simply correspond to Clifford gates on all $\hat{\imath}$ qubits. The remaining operation $\tilde{U} = (c_1 I + c_2 D_{\hat{d}_2}S_{\hat{s}_2} S_{\hat{s}_1}D_{\hat{d}_1})$ can then be written as a multi-qubit rotation operation
\begin{align}
    \tilde{U} &= \cos(\theta) I + e^{i\alpha} \sin(\theta) D_{\hat{d}_2}S_{\hat{s}_2} S_{\hat{s}_1}D_{\hat{d}_1} \\
    &= \cos(\theta) I + e^{i\alpha} (-1)^{\hat{d}_2 \cdot (\hat{s}_1+\hat{s}_2)}\sin(\theta) D_{\hat{d}_2}D_{\hat{d}_1} S_{\hat{s}_2}S_{\hat{s}_1} \label{eq:U}
\end{align}
where one can find that $c_1 = \cos(\theta)$ and $c_2 = e^{i\alpha}\sin(\theta)$ by computing $U U^\dagger$ from Equation \eqref{eq:udecomp} (see Annex A, Lemma 2 of~\cite{masot-llima_stabilizer_2024}) and where we utilise the (de)stabilizer commutation relation. One can readily see that this is of the form of a multi-qubit rotation operation over the qubits $\hat{s}_1+\hat{s}_2$ and $\hat{d}_1+\hat{d}_2$. 

For multi-qubit operators that have more than two terms in their decomposition, one may recursively apply the above procedure between the $1$\textsuperscript{st} to $2..i$\textsuperscript{th} elements of the decomposition to determine a nested sequence of controlled rotations. However given that one typically constructs multi qubit operations given a basis of Cliffords operators and arbitrary $Z$ rotations, which only require two terms in their decomposition, we do not further consider multi-qubit non-Clifford operations. 
\subsection{Expectation Values}
To determine expectation values of some operator $O$ one must determine the decomposition in the form of Equation \eqref{eq:udecomp}, noting that for Pauli expectation values this can always be done with a single term, i.e. $O=\alpha D_{\hat{a}}S_{\hat{b}}$. Given this one can find the expectation value on the MPS directly by first applying $Z\ (X)$ gates that result from the $D_{\hat{a}}$ ($S_{\hat{b}}$) then following the standard prescription for computing expectation values~\cite{orus2014practical}. We note that provided that the MPS is in canonical form, the computation of local expectation values may be performed in polynomial time. 
\subsection{Projection} \label{apn:stn_proj}
Given an expectation value, as determined above, one may perform a projective measurement by performing the computation $\frac{1+pO}{2}\ket{\psi}$ where $p$ is the measurement outcome selected based off $\expval{O}$. This results in the following outcome,
\begin{equation}
    \frac{1+pO}{2}\ket{\psi} = \frac{1}{2}\sum_{i=1}^{2^N} c_i(D_{\hat{\imath}}+\alpha p (-1)^{\hat{a}\cdot \hat{\imath}}D_{\hat{\imath}\cdot \hat{b}})\ket{\phi}. \label{eq:proj}
\end{equation}
This may be performed using the following operation on the MPS
\begin{equation}
    \text{Proj}_p = \ket{0}\!\bra{0}_k \frac{1}{\sqrt{2}}I \pm \frac{\alpha (-i)^{\hat{a}\cdot \hat{b}}}{\sqrt{2}} D_{\hat{n}} D_{\hat{a}} D_{\hat{b}} \label{eq:proj_op}
\end{equation}
where $k$ corresponds to the first destabilizer row that anti-commutes with $O$. One must then perform the projective measurement on the stabilizer tableau by following the standard formalism as outlined in~\cite{aaronson2004improved} and re-normalise the MPS.

An alternative way to view projection is purely via the stabilizer tableau, where we note that the procedure to update the stabilizer tableau results in the multiplication of (de)stabilizer rows that anti-commute with the operator, which is precisely what is occurring in Equation \eqref{eq:proj}. The tableau update procedure also results in the projection of the first destabilizer row that anti-commutes with $O$, which again directly corresponds to what is occurring in the MPS.

An interesting observation to make here, and one that is pertinent in this work is that projection is a computationally complex operation as the operator in Equation \eqref{eq:proj_op} is an operation akin to a controlled rotation,
similar to that in Equation~\eqref{eq:U} for non-Clifford operations. 

\section{Analysing the computational complexity of MAST}\label{app:proof}
The cost of MAST simulation results from measurement projection, this can be of either the magic register \emph{or} the data register. The tensor-network has the form $\ket{\nu} = \ket{0}^{\otimes N} \prod_{i}^{t} R_X(\theta_i)\ket{0}$. The projection operation for an operator $O = \alpha D_{\hat{a}} S_{\hat{b}}$ as per~\cite{masot-llima_stabilizer_2024} is
\begin{align}
    \frac{I+O}{2}&=\frac{1}{\sqrt{2}}\ket{0}\bra{0}_k(I \pm \alpha(-i)^{\hat{a} \cdot \hat{b}} X_{I_X}Y_{I_Y}Z_{I_Z}) \nonumber \\
    &= \ket{0}\bra{0}_k \mathcal{R}
\end{align}
where $k$ is the first non-zero element of $\hat{a}$, $I_Y=\hat{a}+\hat{b} \mod{2}$, $I_X=\hat{a}+I_Y \mod{2}$ and $I_Z=\hat{b}+I_Y\mod{2}$. This operation can be rewritten as
\begin{align}
    \frac{I+O}{2} &= \frac{1}{4\sqrt{2}}(I+Z_k)\mathcal{R} \nonumber \\
    &= \frac{1}{4\sqrt{2}}(\mathcal{R}  + (Z_k \pm  \alpha(-i)^{\hat{a}\cdot\hat{b}} Z_kX_{I_X}Y_{I_Y}Z_{I_Z})) \nonumber \\
    &= \frac{1}{4\sqrt{2}}(\mathcal{R}  + (Z_k \mp  \alpha(-i)^{\hat{a}\cdot\hat{b}} X_{I_X}Y_{I_Y}Z_{I_Z}Z_k)) \label{eq:proj}
\end{align}
where we have used that $k$ is either in $I_X$ or $I_Y$, but not both, and hence will anti-commute through $X_{I_X} Y_{I_Y} Z_{I_Z}$. There are two notable regions of interest here, one where $k<N$ and one where $k\geq N$. Applying Equation~\eqref{eq:proj} to the former case results in 
\begin{align*}
    \frac{I+O}{2}\ket{\nu} &= \frac{1}{4\sqrt{2}}(\mathcal{R}\ket{\nu} + (Z_k\ket{\nu} \\
    &\qquad \mp  \alpha(-i)^{\hat{a}\cdot\hat{b}} X_{I_X}Y_{I_Y}Z_{I_Z}Z_k)\ket{\nu}) \\
    &= \frac{1}{4\sqrt{2}}(\mathcal{R}\ket{\nu} + (\ket{\nu} \mp  \alpha(-i)^{\hat{a}\cdot\hat{b}} X_{I_X}Y_{I_Y}Z_{I_Z})\ket{\nu}) \\
    &= \frac{1}{4\sqrt{2}} (\ket{\nu} \pm  \alpha(-i)^{\hat{a}\cdot\hat{b}} X_{I_X}Y_{I_Y}Z_{I_Z}\ket{\nu}+\ket{\nu}) \\
    &\qquad \mp \alpha(-i)^{\hat{a}\cdot\hat{b}} X_{I_X}Y_{I_Y}Z_{I_Z}\ket{\nu}) \\
    &=\frac{1}{2\sqrt{2}}\ket{\nu}
\end{align*}
where we recognise that qubit $k$ is in the $\ket{0}^{\otimes N}$ register and is hence unaffected by the $Z_k$ operation. In the regime where $k\geq N$ this simplification cannot be made in general and hence the projection operation \emph{may} be entangling. 

We can however recognise that in the $k>N$ case the only operations on the $\ket{0}^{\otimes N}$ register are $Z_{I_Y}$, hence we can conclude that the data-register is completely unaffected by the projection operation which bounds the bond dimension by $2^{t/2}$, however one finds a significantly tighter bound by considering that there are only $N$ non-trivial stabilizer rows, and hence the logical projection operation occurs over at most $N$ qubits, one of which is projected in the tensor network, with any further logical projection occurring over a subset of the same $N$ qubits, except for the one that is projected, with an additional unentangled qubit now included. This means that at most there can be only $N$ entangled qubits after a logical projection, which bounds the bond dimension by $2^{N/2}$. 

Given this, one may consider the computational complexity of MAST to be directly related to the location of the first destabilizer term in the projection operator. However, this is only an upper-bound, but as we will see this result is all that is needed to demonstrate the scaling observed for $T$-doped Clifford circuits seen in the main text.  

\subsection{The operator decomposition for T-doped Cliffords}
\begin{figure}
    \centering
    \begin{tikzpicture}
        \matrix [matrix of nodes,nodes={minimum size=2cm, anchor=center},column sep=1mm, row sep=1mm, left delimiter={[}, right delimiter={]}, nodes in empty cells, ampersand replacement=\&] (m) {
          \node[fill=red!20] {Random}; \& \node[fill=blue!20] {Random($X$)}; \\
          \node[fill=yellow!20] {$I^{\otimes N}$}; \& \node[fill=green!20] {diag($X$)}; \\
        };
        \draw[decorate,decoration={brace,amplitude=10pt}, transform canvas={xshift=-1.4em}, thick] (m.west) -- (m.north west) node[midway,xshift=-10pt,left] {$N$};
        \draw[decorate,decoration={brace,amplitude=10pt}, transform canvas={yshift=0.5em}, thick] (m.north west) -- (m.north) node[midway,yshift=10pt,above] {$N$};
        \draw[decorate,decoration={brace,amplitude=10pt}, transform canvas={xshift=-1.4em}, thick] (m.south west) -- (m.west) node[midway,xshift=-10pt,left] {$t$};
        \draw[decorate,decoration={brace,amplitude=10pt}, transform canvas={yshift=0.5em}, thick] (m.north) -- (m.north east) node[midway,yshift=10pt,above] {$t$};
    \end{tikzpicture}
    \caption{The structure of the stabilizer tableau in MAST after executing a random circuit, but before projecting the register of magic state qubits. It consists of four regions, demarcated by the data and magic state registers. For our purposes, the right half of the tableau (i.e. the columns corresponding to the magic state qubits) are of interest. The lower right quadrant has diagonal $X$ terms, with $I$ terms on the off-diagonal. In the upper right quadrant, there is a random distribution of $X$ terms. The upper left quadrant of the table is a random tableau and not relevant to this discussion, while each row of the lower left quadrant is the identity Pauli string (i.e. $I^{\otimes N}$. This form may be found by observing that operations on the data register only modify the left half of the tableau, while magic state injection operations only add $X$ terms to the right half of the tableau.}
    \label{fig:random_destab_layout}
\end{figure}

\begin{figure}[t!]
    \setlength\figureheight{0.75\linewidth}
    \setlength\figurewidth{0.95\linewidth}
    \centering
\begin{tikzpicture}

\definecolor{darkgrey176}{RGB}{176,176,176}
\definecolor{darkorange25512714}{RGB}{255,127,14}
\definecolor{lightgrey204}{RGB}{204,204,204}
\definecolor{steelblue31119180}{RGB}{31,119,180}

\begin{axis}[
height=\figureheight,
legend cell align={left},
legend style={
  fill opacity=0.8,
  draw opacity=1,
  text opacity=1,
  at={(0.03,0.97)},
  anchor=north west,
  draw=lightgrey204
},
tick align=outside,
tick pos=left,
width=\figurewidth,
x grid style={darkgrey176},
xlabel={Layers \(\displaystyle (L)\)},
xmin=0.0499999999999999, xmax=20.95,
xtick style={color=black},
y grid style={darkgrey176},
ylabel={\(\displaystyle 2\log(\chi)/N\)},
ymin=0.0970511825650751, ymax=0.161925166133423,
ytick style={color=black}
]
\addplot [semithick, steelblue31119180, mark=+, mark size=3, mark options={solid}, only marks]
table {%
1 0.1
2 0.100000137586055
3 0.100000412757772
4 0.10000096309963
5 0.100002063777049
6 0.100004265106695
7 0.100008667665223
8 0.100017472379251
9 0.100035080195426
10 0.100070289382086
11 0.100140681987325
12 0.100281364242718
13 0.100562317868557
14 0.101122589022699
15 0.102236645833371
16 0.104439278518939
17 0.108746154632156
18 0.116992377845405
19 0.132192699419798
20 0.158496158348006
};
\addlegendentry{\(\sum_{i}^{L} 2/2^{N-i}\)}
\addplot [semithick, darkorange25512714, mark=x, mark size=4, mark options={solid, thick}, only marks]
table {%
1 0.1
10 0.100144197417391
11 0.100288250853312
12 0.100144197417391
13 0.10100636833447
14 0.101435529297707
15 0.102290040211008
16 0.104264433740849
17 0.108814159686902
18 0.11583370271671
19 0.132423456227053
2 0.1
20 0.158976348698498
3 0.1
4 0.1
5 0.1
6 0.1
7 0.1
8 0.1
9 0.100144197417391
};
\addlegendentry{MAST}
\end{axis}

\end{tikzpicture}
    \caption{The expected bond-dimension for an $N=20$ $T$-doped Clifford circuit (in blue) along with the simulated bond-dimension using MAST (in orange). We find that the expected result is very close to the findings from the simulation. We note that our expression for the expected bond-dimension accounts only for a single increase in bond-dimension to $\chi=4$ as that is the primary mechanism by which bond-dimension increases for $t<N$.}
    \label{fig:stats}
\end{figure}
\begin{figure*}
    \begin{tikzpicture}
    \node at (0, 0) {
        \begin{quantikz}[row sep=6mm]
            \qw & \ctrl{4} & \qw \\
            \qw & \control{} & \qw \\
            \\ \\
            \qw & \targ & \qw & \qw
        \end{quantikz} 
         =
         \begin{quantikz}[row sep=2mm]
             \qw & \qw & \ctrl{3} & \targ{} & \qw & \gate{T^\dagger} & \qw & \targ{} & \ctrl{3} & \qw & \qw & \qw & \qw & \ctrl{1} & \qw \\
             \qw & \qw & \qw & \targ{} & \ctrl{2} & \gate{T^\dagger} & \ctrl{2} & \targ{} & \qw & \qw & \qw & \qw & \qw & \control{} & \qw \\
             \lstick{$\ket{0}$} & \gate{H} & \qw & \ctrl{-2} & \qw & \gate{T} & \qw & \ctrl{-2} & \qw & \gate{H} & \gate{S} & \ctrl{2} & \gate{H} & \meter{} \wire[u][1]{c} \\
             \lstick{$\ket{0}$} & \qw & \targ{}  & \qw & \targ{} & \gate{T} & \targ{} & \qw & \targ{} & \rstick{$\ket{0}$} \\
             \qw & \qw & \qw & \qw & \qw & \qw & \qw & \qw & \qw & \qw & \qw & \targ{} & \qw & \qw & \qw
         \end{quantikz}
    };
    \end{tikzpicture}
    \caption{A Toffoli gate ($CCX$) decomposition as per Ref.~\cite{bravyi2016improved} with $4$ non-Clifford operations ($T$, $T^\dagger$) and $8$ two-qubit Clifford operations. This decomposition requires $2$ ancillary qubits, one of which may be reused. This may be turned into a decomposition for $CCZ$ by application of $H$ gates at the start and end of the target register.}
    \label{fig:ccz_ancilla}
\end{figure*}
\begin{figure*}
    \begin{tikzpicture}
    \node at (0, 0) {
        \begin{quantikz}[row sep=6mm]
            \qw & \ctrl{2} & \qw \\
            \qw & \control{} & \qw \\
            \qw & \targ & \qw & \qw
        \end{quantikz} 
         =
         \begin{quantikz}[row sep=2mm]
            \qw & \qw & \qw & \qw & \ctrl{2} & \qw & \qw & \qw & \ctrl{2} & \qw & \ctrl{1} & \gate{T} & \ctrl{1} & \qw \\
            \qw & \qw & \ctrl{1} & \qw & \qw & \qw & \ctrl{1} & \qw & \qw & \gate{T} & \targ{} & \gate{T^\dagger} & \targ{} & \qw \\
            \qw & \gate{H} & \targ{} & \gate{T^\dagger} & \targ{} & \gate{T} & \targ{} & \gate{T^\dagger} & \targ{} & \gate{T} & \gate{H} & \qw & \qw & \qw \\
         \end{quantikz}
    };
    \end{tikzpicture}
    \caption{A standard Toffoli gate ($CCX$)  decomposition~\cite{nielsen2000quantum} with $7$ non-Clifford operations ($T$, $T^\dagger$) and $6$ two-qubit Clifford operations. Note like Figure~\ref{fig:ccz_ancilla}, one may turn this decomposition into one for the $CCZ$ gate by application of two $H$ gates.}
    \label{fig:ccz_no_ancilla}
\end{figure*}
\begin{figure}
    \hspace{-5mm}
    \includegraphics{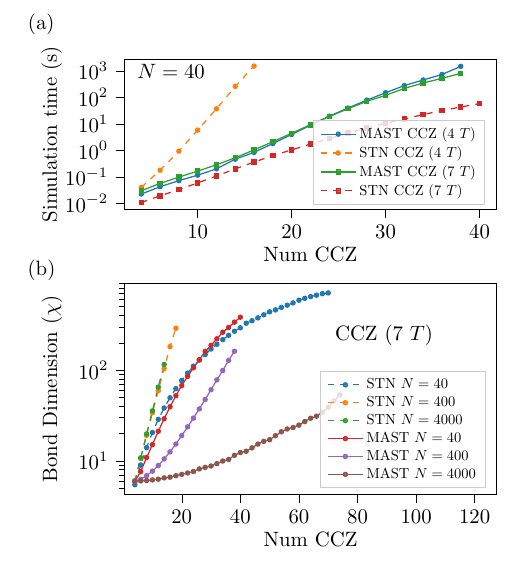}
    \caption{(a) Simulation cost of the Hidden Bit Shift circuit on $40$ qubits. Two cases are simulated, one where the CCZ gates are decomposed into $4$ $T$-gates + two ancillas, as per~\cite{bravyi2016improved}, and one where they are decomposed into $4$ $T$-gates with no extra ancillas. Details of these two decompositions can be found in Appendix~\ref{apn:ccz}. While conventional STNs are dependent on the decomposition used, MAST is indifferent. (b) Larger simulations using the $7$ $T$ decomposition. Simulations were run on Dual 2.45GHz AMD EPYC processors. Each point is an average over 10,000 shots.}\label{fig:hidden-shift-end}
\end{figure}
For the $T$-doped Cliffords found in the main body of this work, the computational complexity arises purely from the projection of the magic register, and as such we will restrict ourselves to analysis of this regime only. The operator we seek to project is $O=Z_i$ where $N\leq i < N+t$. To find the operator decomposition we first find all rows of the stabilizer and destabilizer tableaus that anti-commute with the operator $O$, and note that the first such stabilizer row $S_k$ corresponds to the $k$ defined above. We recognise that for a random $T$-doped Clifford circuit, the tableau will have the form shown in Figure~\ref{fig:random_destab_layout} before any qubits are projected (we describe below how this tableau will change during projection). For each column in the upper right block, we can establish that the probability of there being an $X$ (and hence anticommuting with $O$) approaches $0.5$ (discussed further below). Given this, the probability of at least one of the columns containing an $X$ in the data register rows of the stabilizer tableau is asymptotically $1 - \frac{1}{2^{N-w}}$ where $w$ is the number of qubits projected thus far. After a given decomposition is found and the projection operation is applied to the tensor network, we also project the $k$th qubit in the stabilizer tableau. This may result in stabilizer rows being multiplied with the row $S_k$ we selected earlier, however we note that the product of two random binary strings is another random binary string, which does not alter the statistics of $X$s on the magic register columns (on the rows that haven't been projected yet). The final step of projection on the stabilizer tableau is to replace the $k$th destabilizer row with the $k$th stabilizer, and set the $k$th stabilizer row to the operator being projected. Hence, the process of projecting the magic register results in a reduction of the number of stabilizer rows in the top half of the tableau which may have an $X$ in column $i'$ at the next round of projection. We plot the statistics of magic-register measurement in the $t<N$ regime by assuming that the only process in which the bond-dimension increases is by going from $\chi=2$ to $\chi=4$. This is shown in Figure~\ref{fig:stats}, where we find a very close resemblance to the scaling seen in Figure~\ref{fig:rand}. We reiterate however that this model does not account for simplifications that may be made by compression of the tensor network for specific instances, which may have a significant affect for more structured circuits.

We can also extend this analysis to measurement in the data register, where we recognise that after the projection of the magic register, there are $N-t$ non-trivial stabilizer rows that are in data register with each column having an $X$ with probability $0.25$, and $Y$ with probability $0.25$, hence as a result the chance of selecting a data register row is asyptotically $1-\frac{1}{2^{N-t-w}}$ for $t<N$ and $w$ measured qubits. This suggests that one may measure up to $N-t$ qubits efficiently in the $t<N$ regime. 

\subsection{Probability of $S_{i, j}$ containing an $X$}
Here we argue that the probability $p_{i,j}(n)$ that a stabilizer tableau element $S_{i,j}$ of a uniformly random sampled Clifford is $X$ or $Y$ is $\frac{2^{n-1}}{2^n - 1}$.

We begin by noting that an arbitrary element of the Clifford group is equivalent to an element from the symplectic group ${\rm Sp}(2n, \mbf_2)$ and an element from the Pauli group~\cite{10.1063/1.4903507}. The matrix elements of the stabilizer tableau $S_{i, j}$ can be directly read from the matrix elements of the corresponding element in the symplectic group~\cite{10.1063/1.4903507}. We further note that as the row (and indeed column) permuting operators are themselves elements of the Clifford group (and so the uniform measure on the (finite) Clifford group is invariant under their action), $p_{i,j}(n)$ is independent of $i$ and $j$. Let us therefore consider -- without loss of generality -- the probability that the top-left element is one. The top row of $M$ is uniformly distributed across the non-zero vectors in $\mbf_2^n$ (by the transitivity of the action of ${\rm Sp}(n;\mbf_2)$ on that space); we therefore immediately see that the probability of the first element being one is 
\begin{equation}
   p(n)= \frac{|\{v\in \mbf_2^n \backslash \{\boldsymbol{0}\} : v_1=1\}|}{|\mbf_2^n \backslash \{\boldsymbol{0}\}|}=\frac{2^{n-1}}{2^n-1},
\end{equation}
as desired.

\section{$CCZ$ Gate Decompositions} \label{apn:ccz}
The main body of this work utilised the decomposition of the $CCZ$ gate used in Ref.~\cite{bravyi2016improved,bravyi2019simulation} as shown in Figure~\ref{fig:ccz_ancilla}. The primary motivation for this decomposition is to minimise the number of non-Clifford operations required as that is the expensive resource for stabilizer simulators. The trade-off here is that this decomposition requires ancilla qubits, and additionally requires more entangling gates compared to the standard decomposition as shown in Figure~\ref{fig:ccz_no_ancilla}. Hence the decomposition in Figure~\ref{fig:ccz_ancilla} is suited to simulation methods constrained by magic, whereas the decomposition in Figure~\ref{fig:ccz_no_ancilla} is well suited to simulation methods constrained by entanglement.

\section{Hidden Bit Shift}\label{app:hidden-shift}
Simulations of the Hidden Bit Shift circuit~\cite{bravyi2016improved,bravyi2019simulation} typically use a four $T$-gate decomposition of the $CCZ$ gate, as these techniques scale only with the number of non-Clifford operations. However, this requires the addition of two ancilla qubits. To explore the effect of this, we also simulated identical Hidden Bit Shift circuits with the standard 7 $T$-gate decomposition of the Toffoli gate~\cite{nielsen2000quantum}. Details of these can be found in Appendix~\ref{apn:ccz}. Figure~\ref{fig:hidden-shift-end} shows that with the $7$ $T$-gate decomposition, the simulation time for STN is lower compared to the $4$ $T$-gate decomposition, while it remains unchanged for MAST despite the circuit containing significantly more $T$-gates. These results are explained by the ancilla qubits and extra entangling gates present in the $4$ $T$ CCZ, which increase the simulation time for STN. For larger systems shown in Figure~\ref{fig:hidden-shift-end}(b) the constant factor advantage of requiring fewer ancillas is drowned out by the exponential cost of magic and entangling gates.
\end{document}